\documentclass[twocolumn,aps,superscriptaddress,prb,nofootinbib,longbibliography,preprintnumbers]{revtex4-2}
\usepackage{graphicx} 
\usepackage{amsmath,amssymb}
\usepackage{slashed}
\usepackage{hyperref}
\usepackage[dvipsnames]{xcolor}
\usepackage{multirow}
\usepackage{bbm}
\usepackage[english]{babel}
\usepackage{natbib}
\usepackage{ulem}
\setcitestyle{numbers,square}
\usepackage{newtxmath}  
\usepackage{slashed}
\usepackage{tikz}
\usetikzlibrary{decorations.markings,decorations.pathmorphing}

\def\tr{\mathrm{tr}}

\def\={\stackrel{\bullet}{=}}

\def\({\left(}
\def\){\right)}
\def\[{\left[}
\def\]{\right]}

\def \be {\begin{equation}}
\def \ee {\end{equation}}
\def \beqa {\begin{eqnarray}}
\def \eeqa {\end{eqnarray}}
\def \beal#1 {\begin{align}#1\end{align}}
\def \bes#1 {\begin{equation}\begin{split}#1\end{split}\end{equation}}

\def\x't{(\boldsymbol{x'},t)}

\def\3tensor#1#2#3#4{#1^{#2\;#4}_{\;\;#3}}

\def\i{\mathrm{i}}

\def\ccr{c_{\boldsymbol{r}}}

\def\ar{a_{\boldsymbol{r}}}
\def\br{b_{\boldsymbol{r}}}

\begin{document}
\preprint{OU-HET-1287}
\title{Non-singlet conserved charges and anomalies in 3+1~D staggered fermions }
\author{Tetsuya Onogi}
\email[]{onogi@het.phys.sci.osaka-u.ac.jp}
\affiliation{Department of Physics, The University of Osaka, Toyonaka, Osaka 560-0043, Japan}

\author{Tatsuya Yamaoka}
\email[]{t\_yamaoka@het.phys.sci.osaka-u.ac.jp}
\affiliation{Department of Physics, The University of Osaka, Toyonaka, Osaka 560-0043, Japan}


\begin{abstract}
In this paper, we show that the 3+1~D staggered fermion Hamiltonian possesses, in addition to the conserved charge $Q_0$ that generates the vector $\mathrm{U}(1)_V$ transformation, conserved charges $Q_F$ that generate the $\mathrm{SU}(2)_A$ transformations in the continuum limit, acting simultaneously on left- and right-handed Weyl fermions in opposite directions. Each conserved charge $Q_F$ can be regarded as the generator of a $\mathrm{U}(1)_F$ subgroup of $\mathrm{SU}(2)_L \times \mathrm{SU}(2)_R \times \mathrm{U}(1)_A$. One of these lattice charges satisfies the Onsager algebra. 
On the lattice, the charges $Q_F$ do not commute with $Q_0$, and no symmetric mass term exists that commutes with both $Q_0$ and $Q_F$. This signals the presence of a mixed anomaly. Remarkably, however, in the continuum limit, a symmetric mass term commuting with both $Q_0$ and $Q_F$ can be constructed. 
This means that the mixed anomaly that is nontrivial on the lattice becomes trivial in the IR QFT obtained in the continuum limit, which is consistent with the analysis of the Ward--Takahashi (WT) identity on the lattice.
Indeed, by evaluating this identity associated with the $\mathrm{U}(1)_F$ transformation on the lattice, we confirm that $\mathrm{U}(1)_F$ symmetry is exactly conserved.
\end{abstract}
\maketitle
\section{Introduction}
\label{intro}

On the lattice, any fermion system that respects locality, Hermiticity, 
and translational invariance inevitably gives rise to extra degrees of freedom—commonly referred to as doublers—due to the periodicity of the Brillouin zone. 
These doublers can be removed by introducing a Wilson term; however, this comes at the cost of explicitly breaking chiral symmetry.
This fundamental obstruction is encapsulated in the Nielsen--Ninomiya theorem~\cite{Nielsen:1980rz,Nielsen:1981hk}, which remains a major challenge in the lattice construction of chiral gauge theories.
In contrast, in continuum theories with chiral symmetry, quantum effects generate the Adler–Bell–Jackiw (ABJ) anomaly~\cite{Adler:1969er,Bell:1969ts}, which on the lattice arises from the Wilson mass term~\cite{Karsten:1980wd}.

An alternative strategy is to reinterpret the doubler degrees of freedom as physical flavors, 
as implemented in the staggered fermion formulation~\cite{Kogut:1974ag,Banks:1975gq,Susskind:1976jm}.
By splitting the spinor components, one obtains a description in terms of a single-component complex fermion field, thereby avoiding the need for a Wilson term. 
Nonetheless, once the spinor degrees of freedom are recombined on the lattice, Wilson--like terms inevitably reappear, leading to the breaking of the chiral symmetries present in the continuum theory.
This is, again, a manifestation of the Nielsen--Ninomiya theorem.
In this case, the flavor $\mathrm{SU}(N_f)$ symmetry is also broken to the $\mathrm{U}(1)_\epsilon$ symmetry, which forbids bilinear mass terms in Euclidean spacetime~\cite{Karsten:1980wd,Golterman:1984cy}.

Recently, Ref.~\cite{Chatterjee:2024gje} discussed a Hamiltonian formulation of staggered fermions in 1+1~D,
which is ultra-local and possesses a finite-dimensional Hilbert space.
Remarkably, the model features two conserved, integer-valued lattice charges 
that correspond—upon taking the continuum limit—to the vector and axial charges of a massless Dirac fermion with a perturbative anomaly. 
These charges generate local $\mathrm{U}(1)$ symmetries that can be gauged independently,
yet do not commute at finite lattice spacing. Instead, they satisfy an Onsager algebra whose noncommutativity vanishes in the continuum limit,
thereby providing a realization of the chiral anomaly consistent with the Nielsen--Ninomiya theorem.

Furthermore, Ref.~\cite{Xu:2025hfs} elucidated the anomaly-cancellation mechanism associated with these charges.
One of the present authors proposed a construction of chiral fermions based on the eigenvalues of the axial charge~\cite{Yamaoka:2025sdm},
enabling the realization of certain chiral gauge theories—such as the 3-4-5-0 model—via the symmetric mass generation (SMG) mechanism~\cite{Wang:2013yta,Wang:2018ugf,Razamat:2020kyf,Tong:2021phe,Zeng:2022grc,Wang:2022ucy,Wang:2022fzc,Lu:2022qtc,Onogi:2025tev}. 
SMG refers to the mechanism by which interactions, without introducing fermion bilinear terms, 
gap out a system while preserving symmetries.
Analogous conserved charges satisfying the Onsager algebra have also been identified in higher-dimensional systems~\cite{Gioia:2025bhl,Pace:2025rfu}.

The staggered fermion Hamiltonian in 3+1~D has also been analyzed in detail in Refs.~\cite{Susskind:1976jm,Catterall:2025vrx,Li:2024dpq}.
This Hamiltonian possesses time-reversal symmetry $\mathbb{Z}^{\mathcal{T}}_2$ as well as shift symmetries that translate the staggered fermions along each spatial direction.
The corresponding symmetry operators, denoted by $\mathcal{T}$ and $\mathcal{S}_i$, anticommute, $\{ \mathcal{T}, \mathcal{S}_i \} \neq 0$.
This anticommutation signals the presence of a $\mathbb{Z}_2$ mixed ’t~Hooft anomaly between time-reversal and shift symmetries.
As a result, the vacuum of the system cannot be in a trivially gapped phase that preserves both time-reversal and shift symmetries.
However, this mixed anomaly can be canceled by introducing four copies of staggered fermions with 
\begin{align}
    \label{eq:fourfermi-interaction-term}
    G \sum_{\boldsymbol{r}} \left[ c^{1}_{\boldsymbol{r}}c^{2}_{\boldsymbol{r}}c^{3}_{\boldsymbol{r}}c^{4}_{\boldsymbol{r}} + \text{h.c.} \right],
\end{align}
where $G$ is a coupling constant.
In a sufficiently strong-coupling regime, this interaction dynamically generates a fermion mass gap without breaking the preserved symmetries, leading to a trivially gapped vacuum.
This phenomenon is nothing but SMG.

Moreover, staggered fermions correspond to a discretization of K\"ahler–Dirac fermions
and are therefore invariant under the twisted rotation group, 
which is the diagonal subgroup of the flavor and rotational symmetries~\cite{Banks:1982iq,Golterman:1984cy,Kilcup:1986dg,Catterall:2020fep,Butt:2021brl,Catterall:2022ukg,Catterall:2022jky,Catterall:2025vrx}. 
However, due to the discretization, the rotational symmetry is restricted to the cubic symmetry group corresponding to the cubic lattice, 
and accordingly, the allowed flavor rotation angles are constrained to integer multiples of $\frac{\pi}{2}$. 
In particular, the surviving flavor symmetry corresponds to the shift symmetry of the staggered fermion. 
Specifically, in the continuum limit, the generators $\mathcal{S}_i$ of the shift symmetry correspond to $\frac{\pi}{2}$ flavor rotations 
in the $\mathrm{SU}(2)_L \times \mathrm{SU}(2)_R$ symmetry group.

In this paper, we revisit the analysis of the staggered fermion Hamiltonian in $3+1$ dimensions. 
In particular, we identify conserved, integer-valued lattice charges 
$Q_{\hat{x}_i}$ associated with shift symmetries acting on a Majorana fermion. 
We show that these lattice charges generate axial-flavor $\mathrm{SU}(2)_A$ transformations in the continuum limit, 
under which the left- and right-handed Weyl fermions transform in opposite but simultaneous ways. 
More precisely, these lattice charges can be interpreted as generators of a $\mathrm{U}(1)_{F_i}$ subgroup of $\mathrm{SU}(2)_L \times \mathrm{SU}(2)_R \times \mathrm{U}(1)_A$. 
Importantly, the rotation angles of these transformations are no longer restricted to discrete values, 
but can instead take arbitrary continuous values. 
This extends the results of Ref.~\cite{Catterall:2025vrx}, 
where the rotation angles were constrained by the discrete nature of the lattice symmetries. 

Moreover, the non-commutativity on the lattice between the conserved charge operator $Q_0$, which generates the vector $\mathrm{U}(1)_V$ symmetry in the continuum limit, and these lattice charge operators $Q_{\hat{x}_i}$ signals a mixed 't~Hooft anomaly between the symmetry generated by $Q_0$ and those generated by $Q_{\hat{x}_i}$. 
However, in the large-volume limit ($N \to \infty$) and at finite momentum $\boldsymbol{k}$, 
we show that a symmetric mass term commuting with both $Q_0$ and $Q_{\hat{x}_i}$ can be constructed in the continuum QFT emerging from our lattice theory. 
In terms of the definition of anomaly as an obstruction to a symmetric, trivially-gapped phase\cite{Lieb:1961fr,Affleck:1986pq,Oshikawa2000TopologicalAT,Hastings:2003zx,Chang:2018iay,Wen:2018zux,Thorngren:2019iar},
this demonstrates that the non-trivial lattice anomaly becomes trivial in the continuum limit.
It is well known that in vector-like QFTs, non-singlet chiral symmetries are anomaly-free, 
but the fact that their generators can nevertheless be anomalous at the lattice level is particularly intriguing. 
Such a phenomenon is not unexpected: the transition from a lattice model to its IR description as a QFT entails a fundamental change in the underlying degrees of freedom, under which anomalies need not be preserved~\cite{Tu:2025bqf}. 

Furthermore, this observation has important implications for realizing the SMG mechanism in staggered fermion systems. 
For SMG to be consistently implemented, all symmetries of the lattice theory must be anomaly-free on the lattice. 
However, since $Q_{\hat{x}_i}$ does not commute with $Q_0$ and also fails to commute with the interaction term~\eqref{eq:fourfermi-interaction-term}, 
our analysis suggests that, at least in scenarios where SMG is triggered by the interaction term~\eqref{eq:fourfermi-interaction-term}, 
the symmetry generated by $Q_{\hat{x}_i}$ may need to be explicitly broken.

This paper is organized as follows.  
In Sec.~\ref{Sec:Hamiltonian-Lattice}, we briefly review the Hamiltonian of a single staggered fermion.  
In Sec.~\ref{Sec:conserved-charge}, we investigate the conserved charges $Q_0$ and $Q_{\hat{x}_i}$ associated with this Hamiltonian.  
The construction of these charges is reminiscent of the method proposed in Ref.~\cite{Chatterjee:2024gje} based on the Onsager algebra, 
although some of the conserved charges constructed here do not satisfy the Onsager algebra.  
In Sec.~\ref{Sec:axial-flavor-symmetry}, we show that the conserved charge operators $Q_{\hat{x}_i}$ generate axial-flavor $\mathrm{SU}(2)_A$ transformations in the continuum limit. 
We also demonstrate that, while no mass term exists on the lattice that commutes with both $Q_0$ and $Q_{\hat{x}_i}$, such a symmetric mass term can be constructed in the continuum limit.  

Section~\ref{sec:Conclusion} is devoted to our conclusions.  
Although the main text focuses on the conserved charge operators constructed from single-shift operators acting on Majorana fermions, 
it is also possible to construct conserved charges using double- and triple-shift transformations.  
In Appendix~\ref{apsec:other-conservedCharge}, we investigate the action of these additional conserved charge operators on the fermions in the continuum limit.  
We also comment on the conserved charge operator constructed from the triple shift, 
which was conjectured in Ref.~\cite{Catterall:2025vrx} to generate the singlet axial $\mathrm{U}(1)_A$ symmetry, 
but we find that it does not serve as such a generator.
Finally, in Appendix~\ref{apsec:Anomaly}, under the background gauge field of vector $\mathrm{U}(1)_V$ symmetry, we analyze the Ward--Takahashi (WT) identity for $\mathrm{U}(1)_{F_i}$ symmetry generated by $Q_{\hat{x}_i}$.
Specifically, we confirm that
a residual effect of the symmetry breaking on the lattice does not generate at least relevant or marginal operators which contributes to the anomaly in the continuum limit.

\section{The Hamiltonian on the Lattice}
\label{Sec:Hamiltonian-Lattice}

Here, we review quickly the system of a staggered fermion 
in $3+1$ dimensions~\cite{Kogut:1974ag,Banks:1975gq,Susskind:1976jm,Catterall:2025vrx,Li:2024dpq}.
\subsection{Set up}
\label{subsec:setup}
Let us start with the Hamiltonian,
\begin{align}
    \label{eq:staggered-Hamiltonian}
    H = \frac{\i}{2} \sum_{\boldsymbol{r}} \eta_i(\boldsymbol{r}) \, \ccr^\dagger \nabla_i \ccr \ ,
\end{align}
where $\ccr$ denotes a single-component complex fermion satisfying the canonical anticommutation relation
\begin{align}
    \{c_{\boldsymbol{r}},c_{\boldsymbol{r'}}^\dagger\} = \delta_{\boldsymbol{r},\boldsymbol{r'}} \ ,
\end{align}
and the lattice derivative is defined as
\begin{align}
    \nabla_i \ccr = c_{\boldsymbol{r} + \hat{x}_i} - c_{\boldsymbol{r} - \hat{x}_i} \ .
\end{align}
We assume periodic boundary conditions with $N$ sites in the basis vectors directions, respectively, where the $N$ is even.

The complex fermion $\ccr$ can be decomposed into two Majorana fermions $\ar$ and $\br$ satisfying
\begin{align}
    \ar = \ar^\dagger, \quad \br = \br^\dagger, \quad \{\ar, a_{\boldsymbol{r}'}\} = \{\br, b_{\boldsymbol{r}'}\} = 2 \delta_{\boldsymbol{r}, \boldsymbol{r}'} \ ,
\end{align}
as
\begin{align}
    \ccr = \frac{1}{2} \left( \ar + \i \br \right) \ .
\end{align}
Substituting this into Eq.~\eqref{eq:staggered-Hamiltonian}, the Hamiltonian can be written as
\begin{align}
    \label{eq:Majorana-Hamiltonian}
    H &= H_a + H_b \notag \\
      &= \sum_{\boldsymbol{r}} \frac{\i}{8} \eta_i(\boldsymbol{r}) \left( \ar \nabla_i \ar + \br \nabla_i \br \right) \notag \\
      &= \sum_{\boldsymbol{r}} \frac{\i}{4} \eta_i(\boldsymbol{r}) \left( \ar a_{\boldsymbol{r}+\hat{x}_i} + \br b_{\boldsymbol{r}+\hat{x}_i} \right) \, .
\end{align}

The lattice is composed of four sublattices within each unit square, labeled by 
\begin{align}
    \label{eq:sublattices}
    C\(\boldsymbol{r}_e\) &= \(G_{\boldsymbol{r}_e}, \ B_{\boldsymbol{r}_e}, \ Y_{\boldsymbol{r}_e}, \ R_{\boldsymbol{r}_e}, \) \notag \\ 
    &:=(c_{\boldsymbol{r}_G}, c_{\boldsymbol{r}_B}, c_{\boldsymbol{r}_Y}, c_{\boldsymbol{r}_R}) \ , 
\end{align}
where
\begin{align}
    \boldsymbol{r}_e &= \boldsymbol{r}_G=  (2n_x, 2n_y, n_z) \ , \notag \\
    \boldsymbol{r}_B &= \boldsymbol{r}_G + \hat{x} \ , \notag \\
    \boldsymbol{r}_Y &= \boldsymbol{r}_G + \hat{y} \ , \notag  \\
    \boldsymbol{r}_R &= \boldsymbol{r}_G + \hat{x} + \hat{y} \ . \notag
\end{align}
The basis vectors and reciprocal lattice vectors are given by
\begin{align}
    \label{eq:basis-vectors}
    \vec{a}_x &= \begin{pmatrix} 2 \\ 0 \\ 0 \end{pmatrix}, \quad
    \vec{a}_y = \begin{pmatrix} 0 \\ 2 \\ 0 \end{pmatrix}, \quad
    \vec{a}_z = \begin{pmatrix} 0 \\ 0 \\ 1 \end{pmatrix} \ , \\
    \label{eq:reciprocal-lattice}
    \vec{b}_x &= \begin{pmatrix} \pi \\ 0 \\ 0 \end{pmatrix}, \quad
    \vec{b}_y = \begin{pmatrix} 0 \\ \pi \\ 0 \end{pmatrix}, \quad
    \vec{b}_z = \begin{pmatrix} 0 \\ 0 \\ 2\pi \end{pmatrix} \ ,
\end{align}
satisfying the orthonormal condition $\vec{a}_i \cdot \vec{b}_j = 2\pi \delta_{ij}$.

In momentum space, each component of the complex fermion field is expanded as
\begin{align}
    \label{eq:Fouriermode-c-G}
    G_{\boldsymbol{r}_e} &= \frac{2}{\sqrt{N^3}} \sum_{\boldsymbol{k} \in \text{BZ}} e^{\i \boldsymbol{k} \cdot \boldsymbol{r}_e} \tilde{G}_{\boldsymbol{k}} \ , \quad \text{etc.}
\end{align}
with similar expressions for $ B_{\boldsymbol{r}_e}$, $ Y_{\boldsymbol{r}_e}$, and $ R_{\boldsymbol{r}_e}$,
where $\text{BZ}$ denotes the first Brillouin zone defined by Eq.~\eqref{eq:reciprocal-lattice}
and $ \tilde{G}_{\boldsymbol{k}}$ is a Fourier mode satisfying $\{ \tilde{G}_{\boldsymbol{k}} , \tilde{G}^\dagger_{\boldsymbol{k'}} \} =  \delta_{\boldsymbol{k},\boldsymbol{k'}}$.

In the continuum limit, the low-energy modes are localized near the momenta
\begin{align}
    \label{eq:momentum-of-zeromode}
    \boldsymbol{K}_0 = (0,0,0), \quad \boldsymbol{K}_1 = (0,0,\pi) \ ,
\end{align}
indicating that the infrared theory describes two massless Dirac fermions.

\subsection{Two-flavor free, massless Dirac fermions}
\label{subsec:two-flavorDirac}

Here, let us see that the Hamiltonian~\eqref{eq:staggered-Hamiltonian} describes two-flavor free massless Dirac fermions.
We begin by introducing the following matrices constructed from the gamma matrices,
\begin{align}
    \label{eq:alphaI-alphaR}
    \alpha_i &= \gamma_0 \gamma_i \ , \quad \beta = \gamma_0 \ , \notag \\
    \alpha^r &= \alpha_1^{x} \alpha_2^{y} \alpha_3^{z} \ ,
\end{align}
which satisfy the algebraic relations
\begin{align}
    \{ \alpha_i , \alpha_j \} = 2 \delta_{ij} \ , \quad \{ \alpha_i , \beta \} = 0 \ , \quad \alpha_i^\dagger = \alpha_i \ .
\end{align}
Throughout this section, we work in the Weyl representation,
\begin{align}
    \label{eq:gamma-WeylRep}
    \gamma_0 &= \mathbb{I}_{2\times 2} \otimes \sigma_1 = \begin{pmatrix}
        0 & \mathbb{I}_{2\times 2} \\
        \mathbb{I}_{2\times 2} & 0
    \end{pmatrix} \ , \\
    \gamma_i &= \sigma_i \otimes \i \sigma_2 = \begin{pmatrix}
        0 & \sigma_i \\
        - \sigma_i & 0
    \end{pmatrix} \ , \notag
\end{align}
which leads to
\begin{align}
    \gamma_5 &= -\i \gamma_0 \gamma_1 \gamma_2 \gamma_3 = \mathbb{I}_{2\times 2} \otimes \sigma_3 = \begin{pmatrix}
        \mathbb{I}_{2\times 2} & 0 \\
        0 & - \mathbb{I}_{2\times 2}
    \end{pmatrix} \ , \\
    \alpha_i &= - \sigma_i \otimes \sigma_3 = \begin{pmatrix}
        - \sigma_i & 0 \\
        0 & \sigma_i
    \end{pmatrix} \ .
\end{align}

We then define a matrix-valued fermion field from the staggered fermions as follows~\cite{Catterall:2025vrx},
\begin{align}
    \label{eq:Matrix-Fermion}
    \Phi ( \boldsymbol{R}_e ) &= N_0 \sum_{\{ \boldsymbol{b} \}} c_{\boldsymbol{R}_e + \boldsymbol{b}} \, \alpha^{b} \notag \\
    &= \begin{pmatrix}
        \psi_{1+}(\boldsymbol{R}_e) & \psi_{2+}(\boldsymbol{R}_e) & 0 & 0 \\
        0 & 0 & \psi_{3-}(\boldsymbol{R}_e) & \psi_{4-}(\boldsymbol{R}_e)
    \end{pmatrix} \ ,
\end{align}
where $\{\boldsymbol{b}\}$ is a set of $2^3$ vectors with components $b_i = \{0,1\}$, 
$\boldsymbol{R}_e = 2 \boldsymbol{r}$, $N_0$ is a normalization constant, and $\psi_{1,2}$ ($\psi_{3,4}$) are two-component left-handed (right-handed) Weyl fermions.

We now combine the Weyl fermions $\psi_f$ into two four-component Dirac fermions,  
\begin{align}
    \psi_1 (\boldsymbol{R}_e) &= \begin{pmatrix} \psi_{1+}(\boldsymbol{R}_e) \\ \psi_{3-}(\boldsymbol{R}_e) \end{pmatrix} \, , \quad 
    \psi_2 (\boldsymbol{R}_e) = \begin{pmatrix} \psi_{2+}(\boldsymbol{R}_e) \\ \psi_{4-}(\boldsymbol{R}_e) \end{pmatrix} \, ,
\end{align}
which together form a $2 \times 2$ matrix-valued fermion field,
\begin{align}
    \label{eq:2-2-matrix-fermion}
    \Psi (\boldsymbol{R}_e) 
    = \begin{pmatrix} \psi_1(\boldsymbol{R}_e) & \psi_2(\boldsymbol{R}_e) \end{pmatrix} 
    = \begin{pmatrix}
        \psi_{1+} & \psi_{2+} \\
        \psi_{3-} & \psi_{4-}
    \end{pmatrix} \, .
\end{align}
The Hamiltonian~\eqref{eq:staggered-Hamiltonian} can then be written as
\begin{align}
    \label{Hamiltonian-matrix-2-flavor}
    H &= \i \sum_{\boldsymbol{R}_e} \Psi^\dagger(\boldsymbol{R}_e) \left[
        (\alpha_i \otimes \mathbb{I}_{2 \times 2}) \frac{\nabla_i}{2}  
        + (\beta \gamma_5 \otimes \sigma_i^T) \frac{\nabla_i^2}{2} \right] \Psi(\boldsymbol{R}_e) \, ,
\end{align}
where the matrices $\alpha_i$ and $\beta \gamma_5$ act on the spin indices, while $\sigma_i^T$ acts on the flavor indices. The lattice difference operator $\nabla_i$ acting on $\Psi(\boldsymbol{R}_e)$ is defined by Eq.~(B3) in Appendix B.
This describes two flavors of free, massless Dirac fermions in the low-energy region, since the second term in Eq.~\eqref{Hamiltonian-matrix-2-flavor} vanishes in the continuum limit.

In the momentum space, by the straightforward calculations,
we can find that each the Weyl fermion's Fourier mode $\tilde{\psi}_f$ is 
localized around the momentum $\boldsymbol{K}_0$ or $\boldsymbol{K}_1$
given by Eq.~\eqref{eq:momentum-of-zeromode} in the continuum limit as follows,
\begin{align}
    \label{eq:Fouriermode-Contlimit-psi1}
    \tilde{\psi}_{1+}\(\boldsymbol{k}\) &\propto \begin{pmatrix}
        \tilde{G}_{\boldsymbol{K}_1 + \boldsymbol{k}} + \i \tilde{R}_{\boldsymbol{K}_1 + \boldsymbol{k}} \\
        -\tilde{B}_{\boldsymbol{K}_1 + \boldsymbol{k}} - \i \tilde{Y}_{\boldsymbol{K}_1 + \boldsymbol{k}}
    \end{pmatrix} \, , \\
    \label{eq:Fouriermode-Contlimit-psi2}
    \tilde{\psi}_{2+}\(\boldsymbol{k}\) &\propto \begin{pmatrix}
        -\tilde{B}_{\boldsymbol{K}_0 + \boldsymbol{k}} + \i \tilde{Y}_{\boldsymbol{K}_0 + \boldsymbol{k}} \\
        \tilde{G}_{\boldsymbol{K}_0 + \boldsymbol{k}} - \i \tilde{R}_{\boldsymbol{K}_0 + \boldsymbol{k}}
    \end{pmatrix} \, , \\
    \label{eq:Fouriermode-Contlimit-psi3}
    \tilde{\psi}_{3-}\(\boldsymbol{k}\) &\propto \begin{pmatrix}
        \tilde{G}_{\boldsymbol{K}_0 + \boldsymbol{k}} + \i \tilde{R}_{\boldsymbol{K}_0 + \boldsymbol{k}} \\
        \tilde{B}_{\boldsymbol{K}_0 + \boldsymbol{k}} + \i \tilde{Y}_{\boldsymbol{K}_0 + \boldsymbol{k}}
    \end{pmatrix} \, , \\
    \label{eq:Fouriermode-Contlimit-psi4}
    \tilde{\psi}_{4-}\(\boldsymbol{k}\) &\propto \begin{pmatrix}
        \tilde{B}_{\boldsymbol{K}_1 + \boldsymbol{k}} - \i \tilde{Y}_{\boldsymbol{K}_1 + \boldsymbol{k}} \\
        \tilde{G}_{\boldsymbol{K}_1 + \boldsymbol{k}} - \i \tilde{R}_{\boldsymbol{K}_1 + \boldsymbol{k}}
    \end{pmatrix} \, , 
\end{align}
where $\tilde{G}_{\boldsymbol{k}}, \ \tilde{B}_{\boldsymbol{k}} \ , \tilde{Y}_{\boldsymbol{k}}, \ \tilde{R}_{\boldsymbol{k}} $ are the Fourier modes of the sublattices given by Eq.~\eqref{eq:sublattices}.

\section{The conserved charge operators in the staggered fermions system}
\label{Sec:conserved-charge}

In this section, we identify the conserved charges that commute with the staggered fermion Hamiltonian~\eqref{eq:staggered-Hamiltonian}.
A trivial conserved charge is the fermion number operator $Q_0$, defined as
\begin{align}
    \label{eq:vector-charge}
    Q_0 := \sum_{\boldsymbol{r} \in \Lambda} \left(c_{\boldsymbol{r}}^\dagger c_{\boldsymbol{r}} - \frac{1}{2} \right) = \sum_{\boldsymbol{r} \in \Lambda} \frac{\i}{2} a_{\boldsymbol{r}} b_{\boldsymbol{r}} \ ,
\end{align}
which obviously commutes with the Hamiltonian. Its action on the fermion operators is given by
\begin{align}
    \left[ Q_0, c_{\boldsymbol{r}}^\dagger \right] = c_{\boldsymbol{r}}^\dagger \ , \quad
    \left[ Q_0, a_{\boldsymbol{r}} \right] = -\i b_{\boldsymbol{r}} \ , \quad
    \left[ Q_0, b_{\boldsymbol{r}} \right] = \i a_{\boldsymbol{r}} \ .
\end{align}
In the continuum limit, $Q_0$ becomes the conserved vector charge $\mathcal{Q}_0$, which generates the vector  $\mathrm{U}(1)_V$ symmetry.

To construct additional conserved charges, we define a translation operator $T_{\hat{x}_i}^{(b)}$ that acts only on the Majorana fermions $b_{\boldsymbol{r}}$ along the direction $\hat{x}_i$, such that
\begin{align}
    T_{\hat{x}_i}^{(b)} a_{\boldsymbol{r}} \left(T_{\hat{x}_i}^{(b)}\right)^{-1} &= a_{\boldsymbol{r}} \ , \notag \\
    T_{\hat{x}_i}^{(b)} b_{\boldsymbol{r}} \left(T_{\hat{x}_i}^{(b)}\right)^{-1} &= \zeta_i(\boldsymbol{r}) \, b_{\boldsymbol{r} + \hat{x}_i} \ ,
\end{align}
where $\zeta_i(\boldsymbol{r}) := (-1)^{\sum_{k=i+1}^3 x_k}$~\cite{Catterall:2025vrx}. The Hamiltonian is manifestly invariant under this transformation.

This action can be extended to arbitrary lattice directions $\boldsymbol{\chi} = n_x \hat{x} + n_y \hat{y} + n_z \hat{z}$, where $n_i \in \mathbb{Z}$, through a composite operator
\begin{align}
    T_{\boldsymbol{\chi}}^{(b)} a_{\boldsymbol{r}} \left(T_{\boldsymbol{\chi}}^{(b)}\right)^{-1} &= a_{\boldsymbol{r}} \ , \notag \\
    T_{\boldsymbol{\chi}}^{(b)} b_{\boldsymbol{r}} \left(T_{\boldsymbol{\chi}}^{(b)}\right)^{-1} &=
    \left(\zeta_y(\boldsymbol{r})\right)^{n_y} \left(\zeta_x(\boldsymbol{r})\right)^{n_x} \, b_{\boldsymbol{r} + \boldsymbol{\chi}} \ .
\end{align}

Using these translation operators, we define new conserved charge operators by conjugating $Q_0$,
\begin{align}
    \label{eq:Q-chi-operator}
    Q_{\boldsymbol{\chi}} &:= T_{\boldsymbol{\chi}}^{(b)} Q_0 \left(T_{\boldsymbol{\chi}}^{(b)}\right)^{-1} \notag \\
    &= \frac{1}{2} \sum_{\boldsymbol{r} \in \Lambda} \left(\zeta_y(\boldsymbol{r})\right)^{n_y} \left(\zeta_x(\boldsymbol{r})\right)^{n_x}
    \left(c_{\boldsymbol{r}} + c_{\boldsymbol{r}}^\dagger \right)\left(c_{\boldsymbol{r}+\boldsymbol{\chi}} - c_{\boldsymbol{r}+\boldsymbol{\chi}}^\dagger \right) \notag \\
    &= \frac{\i}{2} \sum_{\boldsymbol{r} \in \Lambda} \left(\zeta_y(\boldsymbol{r})\right)^{n_y} \left(\zeta_x(\boldsymbol{r})\right)^{n_x} a_{\boldsymbol{r}} b_{\boldsymbol{r} + \boldsymbol{\chi}} \ .
\end{align}
Their action on the fermions reads
\begin{align}
    \left[ Q_{\boldsymbol{\chi}}, a_{\boldsymbol{r}} \right] &= -\i \left(\zeta_y(\boldsymbol{r})\right)^{n_y} \left(\zeta_x(\boldsymbol{r})\right)^{n_x} b_{\boldsymbol{r} + \boldsymbol{\chi}} \ , \notag \\
    \left[ Q_{\boldsymbol{\chi}}, b_{\boldsymbol{r}} \right] &= \i \left(\zeta_y(\boldsymbol{r} - \boldsymbol{\chi})\right)^{n_y} \left(\zeta_x(\boldsymbol{r} - \boldsymbol{\chi})\right)^{n_x} a_{\boldsymbol{r} - \boldsymbol{\chi}} \ .
\end{align}
Since both $Q_0$ and $T_{\boldsymbol{\chi}}^{(b)}$ commute with the Hamiltonian, the charge operator $Q_{\boldsymbol{\chi}}$ is also conserved,
\begin{align}
    \left[ H, Q_{\boldsymbol{\chi}} \right] = 0 \ .
\end{align}
However, $Q_{\boldsymbol{\chi}}$ does not commute with the original vector charge $Q_0$,
\begin{align}
    \label{eq:noncommutativity-Q0-Qxi}
    \left[ Q_0 , Q_{\boldsymbol{\chi}} \right]
    &= \frac{\i}{2} \sum_{\boldsymbol{r}} \left(\zeta_y(\boldsymbol{r})\right)^{n_y} \left(\zeta_x(\boldsymbol{r})\right)^{n_x}
    \left( a_{\boldsymbol{r}} a_{\boldsymbol{r} + \boldsymbol{\chi}} - b_{\boldsymbol{r}} b_{\boldsymbol{r} + \boldsymbol{\chi}} \right) \notag \\
    &\neq 0 \ .
\end{align}

This construction of lattice conserved charges closely parallels the method introduced in~\cite{Chatterjee:2024gje},
where the resulting operators satisfy the Onsager algebra~\cite{Onsager:1943jn}. 
Indeed, $Q_0$ and $Q_{\hat{z}}$ obey this algebra, while the other charge operators do not,
due to the presence of position-dependent signs $\zeta_i(\boldsymbol{r})$.~\footnote{
    A detailed investigation of the algebraic structure of the conserved charge operators defined in Eq.~\eqref{eq:Q-chi-operator} is left for future work.
}

In what follows, we focus on the conserved charge operators $Q_{\hat{x}_i}$ associated with elementary lattice shifts.  
The non-commutativity in Eq.~\eqref{eq:noncommutativity-Q0-Qxi} implies a mixed 't~Hooft anomaly between the symmetry generated by $Q_0$ (i.e., $\mathrm{U}(1)_V$) and those generated by $Q_{\hat{x}_i}$.  
Indeed, as argued in Appendix~B of Ref.~\cite{Chatterjee:2024gje}, it is impossible to construct a symmetric mass term on the lattice that commutes with both $Q_0$ and $Q_{\hat{x}_i}$.  
The absence of a trivially gapped phase preserving both the $\mathrm{U}(1)_V$ and $Q_{\hat{x}_i}$ symmetries is precisely the consequence—and in fact the very definition—of an anomaly\cite{Chang:2018iay,Wen:2018zux,Thorngren:2019iar,Choi:2021kmx,Tu:2025bqf}.  

However, in the large-volume limit ($N \to \infty$) and at finite momentum $\boldsymbol{k}$, we will show in the next section that a symmetric mass term commuting with both $Q_0$ and $Q_{\hat{x}_i}$ can be constructed in the continuum QFT obtained from our lattice theory.  
This demonstrates that the nontrivial lattice anomaly becomes trivial in the continuum limit.


\section{Physical interpretation of the conserved charges $Q_{\hat{x}_i}$}
\label{Sec:axial-flavor-symmetry}

Let us now clarify the role of the conserved charge operators $Q_{\hat{x}_i}$.
In momentum space, these charges are expressed as follows,
\begin{align}
    \label{eq:conservedCharge-Qx}
    Q_{\hat{x}} &= \sum_{\boldsymbol{k} \in \text{BZ}} \left[ e^{-\i \frac{\boldsymbol{k} \cdot \vec{a}_x}{2}} \left\{ \cos \left( \frac{\boldsymbol{k} \cdot \vec{a}_x}{2} \right) \left( \tilde{G}_{\boldsymbol{k}+\boldsymbol{b}_3}^\dagger \tilde{B}_{\boldsymbol{k}} - \tilde{Y}_{\boldsymbol{k}+\boldsymbol{b}_3}^\dagger \tilde{R}_{\boldsymbol{k}} \right) \right. \right. \notag \\
    &\quad \left. \left. - \i \sin \left( \frac{\boldsymbol{k} \cdot \vec{a}_x}{2} \right) \left( \tilde{G}_{\boldsymbol{k}}^\dagger \tilde{B}_{-\boldsymbol{k}-\boldsymbol{b}_3}^\dagger - \tilde{Y}_{\boldsymbol{k}}^\dagger \tilde{R}_{-\boldsymbol{k}-\boldsymbol{b}_3}^\dagger \right) \right\} + \text{h.c.} \right], \\
    \label{eq:conservedCharge-Qy}
    Q_{\hat{y}} &= \sum_{\boldsymbol{k} \in \text{BZ}} \left[ e^{-\i \frac{\boldsymbol{k} \cdot \vec{a}_y}{2}} \left\{ \cos \left( \frac{\boldsymbol{k} \cdot \vec{a}_y}{2} \right) \left( \tilde{G}_{\boldsymbol{k}+\boldsymbol{b}_3}^\dagger \tilde{Y}_{\boldsymbol{k}} - \tilde{B}_{\boldsymbol{k}+\boldsymbol{b}_3}^\dagger \tilde{R}_{\boldsymbol{k}} \right) \right. \right. \notag \\
    &\quad \left. \left. - \i \sin \left( \frac{\boldsymbol{k} \cdot \vec{a}_y}{2} \right) \left( \tilde{G}_{\boldsymbol{k}}^\dagger \tilde{Y}_{-\boldsymbol{k}-\boldsymbol{b}_3}^\dagger - \tilde{B}_{\boldsymbol{k}}^\dagger \tilde{R}_{-\boldsymbol{k}-\boldsymbol{b}_3}^\dagger \right) \right\} + \text{h.c.} \right], \\
    \label{eq:conservedCharge-Qz}
    Q_{\hat{z}} &= \sum_{\boldsymbol{k} \in \text{BZ}} \left[ \cos(\boldsymbol{k} \cdot \vec{a}_z) \left( \tilde{G}_{\boldsymbol{k}}^\dagger \tilde{G}_{\boldsymbol{k}} + \tilde{B}_{\boldsymbol{k}}^\dagger \tilde{B}_{\boldsymbol{k}} + \tilde{Y}_{\boldsymbol{k}}^\dagger \tilde{Y}_{\boldsymbol{k}} + \tilde{R}_{\boldsymbol{k}}^\dagger \tilde{R}_{\boldsymbol{k}} \right) \right. \notag \\
    &\quad \left. + \left( e^{\i \boldsymbol{k} \cdot \vec{a}_z} \left( \tilde{G}_{-\boldsymbol{k}} \tilde{G}_{\boldsymbol{k}} + \tilde{B}_{-\boldsymbol{k}} \tilde{B}_{\boldsymbol{k}} + \tilde{Y}_{-\boldsymbol{k}} \tilde{Y}_{\boldsymbol{k}} + \tilde{R}_{-\boldsymbol{k}} \tilde{R}_{\boldsymbol{k}} \right) + \text{h.c.} \right) \right],
\end{align}
where $\vec{a}_i$ are the basis vectors defined in Eq.~\eqref{eq:basis-vectors}, and $\boldsymbol{b}_3 = (0, 0, \pi)^T$.

By using Eqs.~\eqref{eq:Fouriermode-Contlimit-psi1}--\eqref{eq:Fouriermode-Contlimit-psi4} together with Eqs.~\eqref{eq:conservedCharge-Qx}--\eqref{eq:conservedCharge-Qz}, we find that in the large-volume limit ($N \to \infty$), and at finite momentum $\boldsymbol{k}$, the conserved charges act on the matrix fermion $\tilde{\Psi}(\boldsymbol{k})$ [defined in Eq.~\eqref{eq:2-2-matrix-fermion}] as
\begin{align}
    \label{eq:SU2A-transform}
    \lim_{N \to \infty}[Q_{\hat{x}_i}, \widetilde{\Psi}(\boldsymbol{k})] = (\gamma_5 \otimes \sigma_i) \, \widetilde{\Psi}(\boldsymbol{k}) \, ,
\end{align}
where $\gamma_5$ and $\sigma_i$ act on the spin indices and on the flavor indices of the fermion, respectively.
These conserved charges generate an axial-flavor $\mathrm{SU}(2)_A$ transformation,
acting oppositely on left- and right-handed Weyl fermions simultaneously.  
Each conserved charge $Q_{\hat{x}_i}$ can be regarded as the generator of a $\mathrm{U}(1)_{F_i}$ subgroup of $\mathrm{SU}(2)_L \times \mathrm{SU}(2)_R \times \mathrm{U}(1)_A$.  

For simplicity and without loss of generality, we focus on $Q_{\hat{z}}$.  
The $\mathrm{U}(1)_{F_3}$ symmetry generated by $Q_{\hat{z}}$ can be regarded as a non-singlet chiral $\mathrm{U}(1)$ symmetry.  
Under $\mathrm{U}(1)_{F_3}$ transformations, the fermion $\Psi$ transforms as
\begin{align}
     \Psi = \begin{pmatrix}
        \psi_{1+} & \psi_{2+} \\
        \psi_{3-} & \psi_{4-}
    \end{pmatrix}
    \;\longrightarrow\;
    \begin{pmatrix}
        \psi_{1+} & -\psi_{2+} \\
        -\psi_{3-} & \psi_{4-}
    \end{pmatrix} \, .
\end{align}
Remarkably, the following flavor mass term,
\begin{align}
    \label{eq:flavor-mass}
    M_{f} := ( \psi_{1+}^\dagger \psi_{4-} +  \psi_{2+}^\dagger \psi_{3-}) + \text{h.c.} \, ,
\end{align}
provides a fermion mass while preserving both $\mathrm{U}(1)_V$ and $\mathrm{U}(1)_{F_3}$ symmetries.\footnote{In this discussion, we do not consider possible anomalies associated with time-reversal or reflection symmetries.}  
In fact, for finite $\boldsymbol{k}$ and $N \to \infty$,
\begin{align}
    \lim_{N \to \infty} [ Q_0, M_f ] \;=\; \lim_{N \to \infty} [ Q_{\hat{x}_3}, M_f ] \;=\; 0 \, .
\end{align}
Anomalies are interpreted as an obstruction to a symmetric, trivially-gapped phase\cite{Lieb:1961fr,Affleck:1986pq,Oshikawa2000TopologicalAT,Hastings:2003zx,Chang:2018iay,Wen:2018zux,Thorngren:2019iar}.
Thus, while the mixed anomaly between $Q_0$ and $Q_{\hat{x}_3}$ is nontrivial on the lattice, it becomes trivial in the continuum QFT obtained in the large-volume limit. 
As supporting evidence, by evaluating the WT identity associated with $\mathrm{U}(1)_{F_3}$ symmetry under the background gauge field of vector $\mathrm{U}(1)_V$ symmetry, we can see that a residual effect of the symmetry breaking on the lattice does not generate at least relevant or marginal operators which is the contribution to the anomaly (see Appendix ~\ref{apsec:Anomaly}).
This is consistent with the fact that non-singlet chiral symmetries are anomaly-free in QFT.  
Such a phenomenon is not unexpected: passing from a lattice model to its IR description as a QFT involves a fundamental change in the nature of the degrees of freedom, under which anomalies need not be preserved~\cite{Tu:2025bqf}.

\section{Conclusion}
\label{sec:Conclusion}
In the continuum limit, the 3+1~D staggered fermion theory describes a system of free, massless two-flavor Dirac fermions.  
Although the classical $\mathrm{SU}(2)_L \times \mathrm{SU}(2)_R \times \mathrm{U}(1)_A$ symmetry exists in this limit, it is explicitly broken on the lattice by terms analogous to the Wilson term—particularly by the second term in the Hamiltonian~\eqref{Hamiltonian-matrix-2-flavor}.  
Nevertheless, we have shown that the lattice theory still possesses conserved charges $Q_{\hat{x}_i}$ that generate the axial-flavor $\mathrm{SU}(2)_A$ transformations defined in Eq.~\eqref{eq:SU2A-transform} in the continuum limit.  

The noncommutativity between $Q_{\hat{x}_i}$ and the vector charge $Q_0$ signals the presence of a mixed anomaly on the lattice.  
Interestingly, however, in the large-volume limit ($N \to \infty$) at finite momentum $\boldsymbol{k}$, the low-energy QFT admits the flavor mass term~\eqref{eq:flavor-mass}, which commutes with both $Q_0$ and $Q_{F_i}$.  
As a result, the vacuum can be trivially gapped.
As supporting evidence, Appendix~\ref{apsec:Anomaly} shows that by evaluating the WT identity associated with $\mathrm{U}(1)_{F_3}$ symmetry under the background gauge field of vector $\mathrm{U}(1)_V$ symmetry, we can see that a residual effect of the symmetry breaking on the lattice does not generate at least relevant or marginal operators which is the contribution to the anomaly. 
This demonstrates that a nontrivial anomaly in the UV lattice theory can become trivial in the IR QFT description.  
The apparent trivialization occurs because passing from a lattice model to its IR description by a QFT involves a fundamental change in the nature of the degrees of freedom.  

This observation is highly significant.  
In particular, when the UV theory is formulated on the lattice, it suggests that any implementation of the SMG mechanism may require breaking either $Q_0$ or $Q_{\hat{x}_i}$ in order for the theory to remain anomaly-free.  
Although we do not explore this issue in detail in the present work, the fact that the conserved charge operators $Q_{\hat{x}_i}$ fail to commute with the four-fermion interaction term~\eqref{eq:fourfermi-interaction-term} indicates that, for the system to become trivially gapped via the SMG mechanism on the lattice, the $\mathrm{U}(1)_{F_i} \subset \mathrm{SU}(2)_L \times \mathrm{SU}(2)_R \times \mathrm{U}(1)_A$ symmetry must necessarily be broken, at least in scenarios where SMG is triggered by the interaction term~\eqref{eq:fourfermi-interaction-term}\cite{Li:2024dpq,Catterall:2025vrx}.

While the operators, $Q_{\hat{x}}$ and $Q_{\hat{y}}$ clearly do not satisfy the Onsager algebra, they may instead obey a different, as yet unidentified algebraic structure.  
Exploring this possibility is an interesting direction for future work.  

In 1+1~D, the axial charge operator that generates the chiral $\mathrm{U}(1)_A$ symmetry can be constructed from conserved charges that satisfy the Onsager algebra.  
However, as discussed in App.~\ref{apsubsec:transformation-Qxyz}, in 3+1~D no such operator has been found to date.  
This discrepancy may be due to the fact that in 1+1~D, the chiral $\mathrm{U}(1)_A$ symmetry is not intrinsically anomalous but only has a mixed anomaly with $\mathrm{U}(1)_V$, whereas in 3+1~D, the chiral $\mathrm{U}(1)_A$ symmetry is intrinsically anomalous.  
A more detailed study of this issue would be of considerable interest for future research.

\acknowledgments
T.Y. sincerely thanks Soichiro Shimamori and Hiroki Wada for the many fruitful discussions, which were instrumental in shaping the direction and progress of this work.
The work of T.O. is supported in part by JSPS KAKENHI Grant Number 23K03387.
The work of T.Y. is supported in part by JST SPRING, Grant Number JP- MJSP2138.

\onecolumngrid
\appendix

\section{Additional Conserved Charges}
\label{apsec:other-conservedCharge}

In Sec.~\ref{Sec:axial-flavor-symmetry}, we focused on the conserved charges $Q_{\hat{x}_i}$ associated with single lattice translations. 
However, as seen in Eq.~\eqref{eq:Q-chi-operator}, the lattice model admits additional conserved charges corresponding to composite shifts.
In this section, we elucidate the role of these remaining conserved charge operators.

\subsection{Transformations Generated by \texorpdfstring{$Q_{\hat{x}_i \hat{x}_j}$}{Q\_\{x\_i x\_j\}}}
\label{apsubsec:transformation-Qxx}

The conserved charges $Q_{\hat{x}_i \hat{x}_j}$ associated with double-shift transformations are given in momentum space as follows,
\begin{align}
    \label{eq:Q-xy}
    Q_{\hat{x}\hat{y}} &= \sum_{\boldsymbol{k}\in \text{BZ}} \Bigg[ 
    \i \sin \left( \frac{\boldsymbol{k} \cdot \hat{\rho}}{2} \right) 
    \left\{ e^{-\i \frac{\boldsymbol{k} \cdot \hat{\rho}}{2}} \tilde{G}_{\boldsymbol{k}}^\dagger \tilde{R}_{\boldsymbol{k}} 
    - e^{-\i \frac{\boldsymbol{k} \cdot \hat{\sigma}}{2}} \tilde{Y}_{\boldsymbol{k}}^\dagger \tilde{B}_{\boldsymbol{k}} \right\} \notag \\
    &\qquad 
    + \cos \left( \frac{\boldsymbol{k} \cdot \hat{\rho}}{2} \right) 
    \left\{ e^{\i \frac{\boldsymbol{k} \cdot \hat{\rho}}{2}} \tilde{R}_{\boldsymbol{k}}^\dagger \tilde{G}_{-\boldsymbol{k}}^\dagger 
    + e^{-\i \frac{\boldsymbol{k} \cdot \hat{\sigma}}{2}} \tilde{Y}_{\boldsymbol{k}}^\dagger \tilde{B}_{-\boldsymbol{k}}^\dagger \right\} 
    + \text{h.c.} \Bigg], \\
    \label{eq:Q-xz}
    Q_{\hat{x}\hat{z}} &= \sum_{\boldsymbol{k} \in \text{BZ}} \Bigg[
    e^{\i \frac{\boldsymbol{k} \cdot \vec{a}_x}{2}} \i \sin \left( \frac{\boldsymbol{k} \cdot (\vec{a}_x + \hat{z})}{2} \right)
    \left( -\tilde{B}_{\boldsymbol{k}}^\dagger \tilde{G}_{\boldsymbol{k}+\boldsymbol{b}_3} 
    + \tilde{R}_{\boldsymbol{k}}^\dagger \tilde{Y}_{\boldsymbol{k}+\boldsymbol{b}_3} \right) \notag \\
    &\qquad 
    + \cos \left( \frac{\boldsymbol{k} \cdot (\vec{a}_x + \hat{z})}{2} \right)
    \left( \tilde{B}_{\boldsymbol{k}}^\dagger \tilde{G}_{-\boldsymbol{k}-\boldsymbol{b}_3}^\dagger 
    - \tilde{R}_{\boldsymbol{k}}^\dagger \tilde{Y}_{-\boldsymbol{k}-\boldsymbol{b}_3}^\dagger \right) 
    + \text{h.c.} \Bigg], \\
    \label{eq:Q-yz}
    Q_{\hat{y}\hat{z}} &= \sum_{\boldsymbol{k} \in \text{BZ}} \Bigg[
    e^{\i \frac{\boldsymbol{k} \cdot \vec{a}_y}{2}} \i \sin \left( \frac{\boldsymbol{k} \cdot (\vec{a}_y + \hat{z})}{2} \right)
    \left( -\tilde{Y}_{\boldsymbol{k}}^\dagger \tilde{G}_{\boldsymbol{k}+\boldsymbol{b}_3} 
    + \tilde{R}_{\boldsymbol{k}}^\dagger \tilde{B}_{\boldsymbol{k}+\boldsymbol{b}_3} \right) \notag \\
    &\qquad 
    + \cos \left( \frac{\boldsymbol{k} \cdot (\vec{a}_y + \hat{z})}{2} \right)
    \left( \tilde{Y}_{\boldsymbol{k}}^\dagger \tilde{G}_{-\boldsymbol{k}-\boldsymbol{b}_3}^\dagger 
    - \tilde{R}_{\boldsymbol{k}}^\dagger \tilde{B}_{-\boldsymbol{k}-\boldsymbol{b}_3}^\dagger \right) 
    + \text{h.c.} \Bigg],
\end{align}
where we define $\hat{\rho} := \vec{a}_x + \vec{a}_y$ and $\hat{\sigma} := \vec{a}_x - \vec{a}_y$.

These charges act on the matrix fermion $\tilde{\Phi}(\boldsymbol{k})$ [defined in Eq.~\eqref{eq:2-2-matrix-fermion}] as follows,
\begin{align}
     \lim_{N \to \infty}\left[ Q_{\hat{x}\hat{y}}, \widetilde{\Psi}(\boldsymbol{k}) \right] &= \left( \sigma_2 \otimes \sigma_1 \otimes \sigma_1 \right) \widetilde{\Psi}^*(-\boldsymbol{k}) \, , \\
     \lim_{N \to \infty}\left[ Q_{\hat{x}\hat{z}}, \widetilde{\Psi}(\boldsymbol{k}) \right] &= \left( \i \sigma_2 \otimes \sigma_1 \otimes \mathbb{I}_{2 \times 2} \right) \widetilde{\Psi}^*(-\boldsymbol{k}) \, , \\
     \lim_{N \to \infty}\left[ Q_{\hat{y}\hat{z}}, \widetilde{\Psi}(\boldsymbol{k}) \right] &= -\left( \sigma_2 \otimes \sigma_1 \otimes \sigma_3 \right) \widetilde{\Psi}^*(-\boldsymbol{k}) \, .
\end{align}

\subsection{Transformation Generated by \texorpdfstring{$Q_{\hat{x} \hat{y} \hat{z}}$}{Q\_\{xyz\}}}
\label{apsubsec:transformation-Qxyz}
The conserved charge $Q_{\hat{x} \hat{y} \hat{z}}$ associated with a triple-shift operation is expressed in momentum space as
\begin{align}
    Q_{\hat{x} \hat{y} \hat{z}} &= \sum_{\boldsymbol{k} \in \text{BZ}} \Bigg[ 
    \i \sin \left( \frac{\boldsymbol{k} \cdot \hat{R}}{2} \right)
    \left\{ e^{-\i \frac{\boldsymbol{k} \cdot \hat{\rho}}{2}} \tilde{G}_{\boldsymbol{k}}^\dagger \tilde{R}_{\boldsymbol{k}} 
    - e^{-\i \frac{\boldsymbol{k} \cdot \hat{\sigma}}{2}} \tilde{Y}_{\boldsymbol{k}}^\dagger \tilde{B}_{\boldsymbol{k}} \right\} \notag \\
    &\qquad 
    + \cos \left( \frac{\boldsymbol{k} \cdot \hat{R}}{2} \right)
    \left\{ e^{\i \frac{\boldsymbol{k} \cdot \hat{\rho}}{2}} \tilde{R}_{\boldsymbol{k}}^\dagger \tilde{G}_{-\boldsymbol{k}}^\dagger 
    + e^{-\i \frac{\boldsymbol{k} \cdot \hat{\sigma}}{2}} \tilde{Y}_{\boldsymbol{k}}^\dagger \tilde{B}_{-\boldsymbol{k}}^\dagger \right\}
    + \text{h.c.} \Bigg],
\end{align}
where $\hat{R} := \hat{x} + \hat{y} + \hat{z}$.

This operator was interpreted in~\cite{Catterall:2025vrx} as the singlet axial charge operator.
However, its action on the matrix fermion is
\begin{align}
\label{Eq:QXYZ}
     \lim_{N \to \infty}\left[ Q_{\hat{x}\hat{y}\hat{z}}, \widetilde{\Psi}(\boldsymbol{k}) \right] = 
    \left( \sigma_2 \otimes \i \sigma_2 \otimes \i \sigma_2 \right) \widetilde{\Psi}^*(-\boldsymbol{k}) \, ,
\end{align}
which clearly differs from the action of the true singlet axial charge operator.
Hence, the conserved charge $Q_{\hat{x}\hat{y}\hat{z}}$ should not be identified with the singlet axial charge.
Note that
$\mathrm{U}(1)_A$
symmetry in 3+1~D is anomalous by itself while the one in 1+1~D is not.
Our result that $Q_{\hat{x}\hat{y}\hat{z}}$ generates the transformation~\eqref{Eq:QXYZ} is in sharp contrast to that in Ref.~\cite{Catterall:2025vrx}.

\section{Mixed anomalies between the $\mathrm{U}(1)_V$ and $\mathrm{U}(1)_F$ symmetries }
\label{apsec:Anomaly}
We have seen that the mixed anomaly between $Q_{\hat{x}_i}$ and $Q_V$ becomes trivial in the continuum limit. 
By explicitly evaluating the Ward--Takahashi (WT) identity on the lattice, we demonstrate that the $\mathrm{U}(1)_F$ symmetry generated by the conserved charge $Q_{\hat{x}_i}$ is anomaly-free in the continuum limit.

As in the evaluation of the chiral anomaly in the Wilson fermion system, if a symmetry were anomalous, a residual effect of the symmetry breaking on the lattice generates at least relevant or marginal operators contributing to the anomaly. 
However, in the case of the $\mathrm{U}(1)_F$ symmetry, we find no such contribution to the anomaly.

\subsection{From Hamiltonian to Lagrangian}
We start with the lattice Hamiltonian of 2-flavor massless fermions ,
\begin{align}
    \label{eq:general-Hamiltonian-Euclidean}
    H &= \sum_{\boldsymbol{R}_e} \Psi^\dagger(\boldsymbol{R}_e) \hat{h}_E \Psi(\boldsymbol{R}_e) \, ,
\end{align}
where
\begin{align}
    \label{eq:hamiltonian-density}
    \hat{h}_E &= \left[
        (\alpha_i \otimes \mathbb{I}_{2 \times 2}) \frac{\nabla_i}{2}  
        + \beta \Gamma_M^i \frac{\nabla_i^2}{2} \right] \, , \\
         \frac{\nabla_i}{2}\Psi(\boldsymbol{R}_e) &= \frac{1}{2}\[ \Psi(\boldsymbol{R}_e + \hat{i} ) - \Psi(\boldsymbol{R}_e - \hat{i} ) \] \, .
\end{align}
and $\hat{i}$ is an unit  lattice vector in each direction.
Here, we have used the following gamma matrices $\gamma_\mu^E$
\begin{align}
    \gamma_0^M = \gamma_0^E \, , \, \i  \gamma_{i}^M = \gamma_{i}^E \, , \, \gamma_5 ^M = \gamma_5^E \, , \, \i\alpha_i ^{\(M\)} = \alpha_i^{\(E\)} \, , 
\end{align}
so that they satisfy $\{ \gamma_\mu ^E , \gamma_\nu ^E \} = 2 \delta_{\mu\nu}$,
and we have omitted the superscript $E$ for simplicity.
$\Gamma_M^i$ is a tensor matrix that is defined as
\begin{align}
    \label{eq:Def-GammaM}
    \Gamma_M^i & := \begin{cases}
        -\mathbb{I} & \text{for two-flavor Wilson fermions} \, , \\
        \i \gamma_5 \otimes \sigma_i ^T  & \text{for a staggered fermion} \, .
    \end{cases}
\end{align}
The second case in Eq.~\eqref{eq:Def-GammaM} is corresponds to the Hamiltonian given by Eq.~\eqref{Hamiltonian-matrix-2-flavor}.

In momentum space, the Hamiltonian~\eqref{eq:general-Hamiltonian-Euclidean} is
\begin{align}
\label{eq:general-Momentum-Hamiltonian-Euclidean}
    H &= \sum_{\boldsymbol{p} \in \text{BZ}} \widetilde{\Psi}^\dagger (\boldsymbol{p}) \i \hat{h}(\boldsymbol{p}) \Psi (\boldsymbol{p}) \, ,
\end{align}
where
\begin{align}
    \label{eq:hamiltonian-density-Momentum}
    \hat{h}(\boldsymbol{p}) & =  (\alpha_i \otimes \mathbb{I}_{2 \times 2}) \sin (p_i) +\i \beta \Gamma_M^i M(p_i) \ ,
\end{align}
and $M(p_i) = 1 - \cos (p_i)$.

The action associated with the Hamiltonian~\eqref{eq:general-Hamiltonian-Euclidean} is given by\footnote{
In Minkowski spacetime, the action associated to the Hamiltonian~\eqref{eq:general-Hamiltonian-Euclidean} is given by
\begin{align}
    S_M = \int dt \sum_{\boldsymbol{R}_e} ( \Psi ^\dagger \i \partial_0 \Psi -  \Psi ^\dagger \hat{h}_E \Psi ) \ ,
\end{align}
that is transformed under the Wick rotation as follow,
\begin{align}
    S_M = -\i \int dt \sum_{\boldsymbol{R}_e} ( -\Psi ^\dagger \partial_0 \Psi -  \Psi ^\dagger \hat{h}_E \Psi ) \ .
\end{align}
Hence, the action in Euclidean spacetime~\eqref{eq:action--matrix-2-flavor-Euclidian} can be defined as
\begin{align}
    S := -iS_M = \int dt \sum_{\boldsymbol{R}_e} ( \Psi ^\dagger \partial_0 \Psi +  \Psi ^\dagger \hat{h}_E \Psi ) \, .
\end{align}
}
\begin{align}
\label{eq:action--matrix-2-flavor-Euclidian}
    S &=  \int dt \sum_{\boldsymbol{R}_e}  {\overline{\Psi}}(t,\boldsymbol{R}_e) \(  \gamma^0 \partial_0 + \gamma^0 \hat{h}_E \) {\Psi} (t,\boldsymbol{R}_e) \, , \\
    &=  \int dp_0 \sum_{\boldsymbol{p}\in \text{BZ}} \widetilde{\overline{\Psi}}(-p) \mathcal{D}(p) \tilde{\Psi} (p) \, ,
\end{align}
where $\tilde{\bar{\Psi}}(-p) := \tilde{\Psi}^\dagger \gamma^0$ and
\begin{align}
    \mathcal{D}(p) := \i \gamma^0 p_0 + \i \gamma^0 \hat{h}(\boldsymbol{p}) \, .
\end{align}
The Dirac operator $\mathcal{D}(p)$ can be deformed as
\begin{align}
    \mathcal{D}(p) = \i \slashed{s}(p) - \Gamma_M^i M(p_i) \, , 
\end{align}
where
\begin{align}
    \slashed{s}(p) &:= \( \gamma^0 p_0 + \gamma^i \sin \(p_i\) \) \, .
\end{align}
Since $\mathcal{D}^\dagger \mathcal{D} = s^2 + M(p_i)^2 $, a fermion propagator is defined as
\begin{align}
    S_F (p) &:= \mathcal{D}^{-1}(p) = \frac{-\i \slashed{s} - (\Gamma_M ^i)^\dagger M(p_i)}{s^2 + M(p_i)^2 } \notag \\
    &= \begin{cases}
         \frac{-\i \slashed{s} + M(p_i)}{s^2 + M(p_i)^2 } & \text{for two-flavor Wilson fermions} \ , \\
          \frac{-\i \slashed{s} + (\i \gamma_5 \otimes \sigma_i ^T) M(p_i)}{s^2 + M(p_i)^2 } & \text{for a staggered fermion} \, . 
    \end{cases}
\end{align}
Here,
\begin{align}
    s^2 &= p_0^2 + \sum_{\boldsymbol{p}\in \text{BZ}} \sin ^2(p_i) \, , \\
    M(p_i)^2 &=\begin{cases}
         \{ \sum_{\boldsymbol{p}\in \text{BZ}} \( 1 - \cos (p_i) \)\}^2 & \text{for two-flavor Wilson fermions} \ , \\  
          \sum_{\boldsymbol{p}\in \text{BZ}} \( 1 - \cos (p_i) \)^2 & \text{for a staggered fermion} \, .  
    \end{cases}
\end{align}

Let us introduce the gauge field of the vector $\mathrm{U}(1)_V$ symmetry. The action is promoted as
\begin{align}
    \label{eq:gauged-action--matrix-2-flavor-Euclidian}
    S_{\text{gauged}} &=  \int dt \sum_{\boldsymbol{R}_e} \left[\overline{\Psi}(R) \gamma^0 (\partial_0 + \i A_0 )\Psi(R) \right. \notag \\
    & + \frac{1}{2} ( \overline{\Psi}(R) \gamma^i U_{R,i} \Psi (R + \hat{i}) - \overline{\Psi}(R + \hat{i}) \gamma^i U_{R,i}^\dagger \Psi(R)  )  \notag \\
    & \left. +\frac{1}{2} ( \overline{\Psi}(R)  U_{R,i} \Gamma_M^i\Psi(R + \hat{i}) +  \overline{\Psi}(R +\hat{i})U_{R,i}^\dagger  \Gamma_M^i \Psi(R) - 2 \overline{\Psi}(R)  \Gamma_M^i \Psi(R) ) \right] \, ,
\end{align}
where we have omitted the subscription $e$ of the coordinate $R_e$. 
From this action, we can obtain the vertex function as
\begin{align}
    V_\mu (p+q) &:= \begin{cases}
        \i \gamma^0  , & \text{for} \,\mu = 0 \\
        \i ( \gamma^i\cos (\frac{p_i + q_i}{2}a) + \i \Gamma_M^i \sin (\frac{p_i + q_i}{2}a)  ) & \text{for} \,\mu = i .
    \end{cases}
\end{align}
Note that the vertex function satisfies the following relations,
\begin{align}
    \frac{\partial}{\partial p_\mu} S_F^{-1} (k-p) \bigg\vert_{p=0} &= \begin{cases}
        - V_0 (2k)  \, & \text{for} \, \mu = 0  , \\
        - V_i (2k) \, & \text{for} \, \mu = i .
    \end{cases} \, , \\
    \frac{\partial}{\partial q_\mu} S_F^{-1} (k+q) \bigg\vert_{q=0} &= \begin{cases}
         V_0 (2k)  \, & \text{for} \, \mu = 0  , \\
         V_i (2k) \, & \text{for} \, \mu = i .
    \end{cases} \, . 
\end{align}

\subsection{Ward--Takahashi (WT) identity}

Under the background gauge field of the vector $\mathrm{U}(1)_V$ symmetry, we now derive the WT identity associated with the $\mathrm{U}(1)_{\Gamma_5}$ transformation as
\begin{align}
    \label{eq:trasf-Gamma-5}
    \Psi\(\boldsymbol{R}\)  &\to e^{\i \theta \Gamma_5}  \Psi\(\boldsymbol{R}\) \, , \notag \\
    \overline{\Psi}\(\boldsymbol{R}\) &\to \overline{\Psi}\(\boldsymbol{R}\) e^{\i \theta \Gamma_5} \, ,
\end{align}
where $\Gamma_5$ is $\gamma_5 \otimes \mathbb{I}_{2\times2}$ or $\gamma_5 \otimes \sigma_j$.
Under the local transformation in Eq.~\eqref{eq:trasf-Gamma-5}, the variation of the action is
\begin{align}
    \label{eq:localtrsf-Gamma5}
    \theta_{\boldsymbol{R}} \delta_{\boldsymbol{R}} S = \i \theta_{\boldsymbol{R}} \[ -\partial_0 J^0 (R)- \nabla_i ^{-} J^i  \(R\) - X (R)   \] \, ,
\end{align}
where
\begin{align}
    J^0 (R) &= \overline{\Psi} (R) \gamma^0 \Gamma_5 \Psi (R)  \, , \\
    J^i (R) &= \frac{1}{2} \bigg\{\overline{\Psi} (R) \gamma^i \Gamma_5 U_{R,i} \Psi(R + \hat{i}) + \overline{\Psi} (R + \hat{i}) \gamma^i \Gamma_5 U_{R,i}^\dagger \Psi(R)   \bigg\} \, , \\
    X(R) &= \frac{1}{2}\nabla_i ^- \bigg\{\overline{\Psi}(R) \Gamma_M^i \Gamma_5 U_{R,i} \Psi(R + \hat{i}) -  \overline{\Psi} (R + \hat{i}) \Gamma_M^i \Gamma_5 U_{R,i}^\dagger \Psi(R)    \bigg\} \notag \\
    & -\frac{1}{2}  \overline{\Psi} (R) \{\Gamma_5,\Gamma_M^i  \} \bigg(U_{R,i}\Psi(R+\hat{i}  ) + U_{R-\hat{i},i}^\dagger \Psi(R-\hat{i})  -  2 \Psi(R)   \bigg) \, .
\end{align}
For an arbitrary fermionic observable $\mathcal{O}(\Psi,\bar{\Psi})$, the WT identity reads
\begin{align}
    \label{eq:general-WT-id}
    \langle  \[  -\partial_0 J^0 (R)- \nabla_i ^{-} J^i  \(R\) - X (R)  \] \mathcal{O} + \i \delta_{\boldsymbol{R}} \mathcal{O}   \rangle = 0 \, ,
\end{align}
which implies that the first and second terms in Eq.~\eqref{eq:general-WT-id} represents the conventional current, whereas the third term encodes at least relevant or marginal operators which is the contribution to possible mixed anomalies between the $\mathrm{U}(1)_V$ and $\mathrm{U}(1)_{\Gamma_5}$ symmetries after taking the continuum limit.

\begin{figure}[h]
\centering
\begin{tikzpicture}[decoration={markings,mark=at position 0.5 with {\arrow{>}}}]
\draw[thick,postaction={decorate}] (0,0) -- (2,1);
\draw[thick,postaction={decorate}] (2,1) -- (2,-1);
\draw[thick,postaction={decorate}] (2,-1) -- (0,0);
\draw[thick,decorate,decoration={snake,amplitude=1.5pt}] (2,1) -- (4,1) node[right]{$A_\nu(\boldsymbol{q})$};
\draw[thick,decorate,decoration={snake,amplitude=1.5pt}] (2,-1) -- (4,-1) node[right]{$A_\mu(\boldsymbol{p})$};
\fill (0,0) circle (4pt);
\node at (-0.5,0) {$X(\boldsymbol{s})$};
\end{tikzpicture}
\caption{Triangle diagram contributing to the anomaly.}
\label{fig:Gamma5-traingle-dig}
\end{figure}
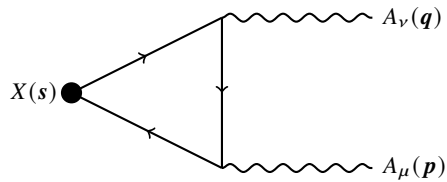
We now evaluate the anomaly contribution from the triangle diagram in Figure~\ref{fig:Gamma5-traingle-dig}, corresponding to the third term in Eq.~\eqref{eq:general-WT-id}.  
In momentum space, $X(R)$ in Eq.~\eqref{eq:general-WT-id} which contributes to the triangle diagram is given by
\begin{align}
    \label{eq:X-momentumspace}
    X(s) &= \widetilde{\overline{\Psi}}(-q) \bigg( \{\Gamma_5 , \Gamma_M^i \} - \big( \Gamma_5 \Gamma_M^i \cos (p_i) + \Gamma_M^i \Gamma_5 \cos (q_i)  \big) \bigg) \widetilde{\Psi}(p) \delta (s - q + p ) \notag \\
    & := \widetilde{\overline{\Psi}}(-q) \hat{V}_5(p,q) \widetilde{\Psi}(p) \delta (s - q + p )  \, , \\
    \hat{V}_5(p,q) &:=   \{\Gamma_5 , \Gamma_M^i \} - \big( \Gamma_5 \Gamma_M^i \cos (p_i) + \Gamma_M^i \Gamma_5 \cos (q_i)  \big)  \, .
\end{align}
For simplicity, we assume temporary gauge fixing, $A_0 = 0$ in the following discussion.
The contribution of the diagram is
\begin{align}
    \label{eq:general-contribution-tri-diagram}
    \Gamma (p+q) &= \Gamma_{\mu,\nu}(pa,qa) A_\mu (p) A_\nu(q) \, , \\
    \Gamma_{\mu\nu}(pa,qa) &= a^2 \int_{\text{BZ}} \frac{d^4k}{(2\pi)^4} \tr [\hat{V}_5 (k +qa,k - pa) S_F ( k - p ) V_\mu (2k - pa ) S_F (k) V_\nu (2k + qa) S_F (k +q) ] \notag \\
    &= a^{-2} \int  \frac{d^4k}{(2\pi)^4} \tr [\hat{V}_5 (pa,qa) S_F ( k - pa ) V_\mu (2k - pa ) S_F (k) V_\nu (2k + qa) S_F (k +q) ] \, ,
\end{align}
where, in the second line, we have defined the dimensionless momentum $k' = {k}a$ and then rewrote ${k}'$ as ${k}$.
Taking the limit $a \to 0$, the following three terms do not vanish,
\begin{align}
    \label{eq:general-nonzero-contribution}
    \Gamma_{\mu\nu}(\boldsymbol{p}a , \boldsymbol{q}a) 
    &\supset \Gamma_{\mu\nu}(0,0) + a\left( p_\alpha \frac{\partial}{\partial (p_\alpha a)} 
            + q_\alpha \frac{\partial}{\partial (q_\alpha a)} \right)  \Gamma_{\mu\nu}(\boldsymbol{p}a , \boldsymbol{q}a) 
       \bigg|_{\boldsymbol{p}a = \boldsymbol{q}a = 0} \notag \\
       & \quad + \frac{a^2}{2} \left( p_\alpha \frac{\partial}{\partial (p_\alpha a)} 
            + q_\alpha \frac{\partial}{\partial (q_\alpha a)} \right) 
       \left( p_\beta \frac{\partial}{\partial (p_\beta a)} 
            + q_\beta \frac{\partial}{\partial (q_\beta a)} \right) 
       \Gamma_{\mu\nu}(\boldsymbol{p}a , \boldsymbol{q}a) 
       \bigg|_{\boldsymbol{p}a = \boldsymbol{q}a = 0} \,   .
\end{align}


\subsubsection{Two-flavor Wilson fermions system}
\label{app-subsubsec:Wilson-fermion-system}
Here, we consider the case where
\begin{align}
    \Gamma_m^i = - \mathbb{I}_{4\times4} \otimes \mathbb{I}_{2\times2} := \Gamma_M\, , \,
    \Gamma_5 = \gamma_5 \otimes \mathbb{I}_{2\times2} \, .
\end{align}
This system corresponds to one of two-flavor Wilson fermions, and the WT identity is associated with the chiral $\mathrm{U}(1)$ symmetry.
 
Since $\{ \Gamma_M ,  \Gamma_5   \} = -2 \gamma_5 \otimes \mathbb{I}_{2\times2}$,
the vertex function $\hat{V}_5$ is given by
\begin{align}
    \hat{V}_5(p,q) &= \gamma_5 \otimes \mathbb{I}_{2\times2} \{ \sum_i^3 ( \cos (p_i) + \cos (q_i)  ) -2  \} \, .
\end{align}

Owing to the trace properties of the gamma matrices, the non-zero contribution from the function~\eqref{eq:general-nonzero-contribution} is given by,
\begin{align}
\label{eq:Wilson-Gamma-munu2} 
\Gamma_{\mu\nu}^{(2)}(\boldsymbol{p}a , \boldsymbol{q}a)  
       & := \frac{a^2}{2} \left( p_\alpha \frac{\partial}{\partial (p_\alpha a)} 
            + q_\alpha \frac{\partial}{\partial (q_\alpha a)} \right) 
       \left( p_\beta \frac{\partial}{\partial (p_\beta a)} 
            + q_\beta \frac{\partial}{\partial (q_\beta a)} \right) 
       \Gamma_{\mu\nu}(\boldsymbol{p}a , \boldsymbol{q}a) 
       \bigg|_{\boldsymbol{p}a = \boldsymbol{q}a = 0} \, \notag \\
       &= -8 p_\alpha q_\beta \int  \frac{d^4k}{(2\pi)^4}  \sum_i^3 M(k_i) \tr \left[ \gamma_5   \frac{\partial}{\partial (p_\alpha a)} S_F ( k - pa ) \, V_\mu (2k - pa ) \, S_F (k) \, V_\nu (2k + qa) \, \frac{\partial}{\partial (q_\beta a)} S_F (k +q) \right] \bigg|_{\boldsymbol{p}a = \boldsymbol{q}a = 0} \notag \\
       & = 8 p_\alpha q_\beta \int  \frac{d^4k}{(2\pi)^4}  \sum_i^3 M(k_i)  \tr \left[ \gamma_5  S_F ( k ) V_\alpha (2k) S_F ( k ) V_\mu (2k) S_F(k) V_\nu (2k) S_F ( k ) V_\beta (2k) S_F ( k ) \right] \notag \\
       &= -32\i  \epsilon^{\mu \nu \alpha\beta}p_\alpha q_\beta \int  \frac{d^4k}{(2\pi)^4}  \sum_i^3 M(k_i) \cos(k_\mu)\cos(k_\nu)\cos(k_\alpha) \frac{\sum_i^3 M(k_i) \cos (k_\beta) - 4 \sin^2(k_{\beta\neq 0}) }{(s^2 + (\sum_i^3 M(k_i))^2)^3 } \, .
\end{align}
This result leads to the existence of the chiral anomaly,
\begin{align}
    \langle X \rangle \propto \epsilon^{\mu \nu \alpha\beta} \partial_\mu A_\nu  \partial_\alpha A_\beta \, ,
\end{align}
where we do not evaluate the integrand on the last line in Eq.~\eqref{eq:Wilson-Gamma-munu2}.

\subsubsection{A single staggered fermion system}
\label{app-subsubsec:staggered-fermion-system}

Here, we consider the case where
\begin{align}
    \Gamma_m^i = \i \gamma_5 \otimes \sigma_i ^T \, , \,
    \Gamma_5 = \gamma_5 \otimes \sigma_3 \, .
\end{align}
Since the anti-commutation relation  between $\Gamma_M^i$ and $\Gamma_5$ is
\begin{align}
    \{ \Gamma_M ^i , \Gamma_5 \} = 2 \i ( \mathbb{I} \otimes \delta_{i3} ) \, ,
\end{align}
$\hat{V}_5$ in Eq.~\eqref{eq:X-momentumspace} is given by
\begin{align}
    \hat{V}_5(p,q) = Y(p_3,q_3) + Z(p_1,p_2,q_1,q_2)\, ,
\end{align}
where
\begin{align}
    \label{eq:Y(pq)}
    Y(p_3,q_3) &:= \i ( \mathbb{I} \otimes \mathbb{I}    )   \{ 2- \cos(p_3) -  \cos(q_3)   \} \, , \\
    \label{eq:Z(pq)}
    Z(\{p_1,p_2\},\{q_1,q_2\}) &:= - \i \sum_{i=1}^2 \{ (\mathbb{I}\otimes \sigma_3\sigma_i^T) \cos(p_i) +  (\mathbb{I}\otimes \sigma_i^T \sigma_3) \cos(q_i)   \}  \, .
\end{align}
Owing to the trace properties of the gamma matrices, it suffices to evaluate the following expression,
\begin{align}
\label{eq:topo-staggered-Gamma-munu2} 
\Gamma_{\mu\nu}^{(2)}(\boldsymbol{p}a , \boldsymbol{q}a)  
       & := \frac{a^2}{2} \left( p_\alpha \frac{\partial}{\partial (p_\alpha a)} 
            + q_\alpha \frac{\partial}{\partial (q_\alpha a)} \right) 
       \left( p_\beta \frac{\partial}{\partial (p_\beta a)} 
            + q_\beta \frac{\partial}{\partial (q_\beta a)} \right) 
       \Gamma_{\mu\nu}(\boldsymbol{p}a , \boldsymbol{q}a) 
       \bigg|_{\boldsymbol{p}a = \boldsymbol{q}a = 0} \notag \\
       &= p_\alpha q_\beta \int  \frac{d^4k}{(2\pi)^4} \, 
          \tr \left[ \hat{V}_5 (k + qa, k - qa) 
          \frac{\partial}{\partial (p_\alpha a)} S_F ( k - pa ) \,
          V_\mu (2k - pa ) \right. \notag \\ 
          & \qquad\qquad\qquad\qquad\qquad\qquad\qquad\qquad\qquad  \times \left. S_F (k) \,
          V_\nu (2k + qa)\,
          \frac{\partial}{\partial (q_\beta a)} S_F (k +q) \right]
          \bigg|_{\boldsymbol{p}a = \boldsymbol{q}a = 0} \, .
\end{align}
After straightforward manipulations, one finds that it vanishes,
\begin{align}
    \Gamma_{\mu\nu}^{(2)}(\boldsymbol{p}a , \boldsymbol{q}a) 
      &= 0 \, ,
\end{align}
since $ Y(k_3+q_3,k_3-p_3)$ given by~\eqref{eq:Y(pq)} yields 
$\Gamma_{\mu\nu}^{(2)}(\boldsymbol{p}a , \boldsymbol{q}a)$ with a flavor trace equal to zero, 
while for $Z(k+qa, k-pa)$~\eqref{eq:Z(pq)} one has
\begin{align}
    Z(\{k_1+q_1a,k_2+q_2a\}, \{k_1-p_1a, k_2-p_2a\}) \bigg|_{\boldsymbol{p}a = \boldsymbol{q}a = 0} = 0 \, .
\end{align}
Thus, a residual effect of the symmetry breaking on the lattice does not generate at least relevant or marginal operators which is the contribution to the anomaly.

\bibliography{reference}

\begin{thebibliography}{41}%
\makeatletter
\providecommand \@ifxundefined [1]{%
 \@ifx{#1\undefined}
}%
\providecommand \@ifnum [1]{%
 \ifnum #1\expandafter \@firstoftwo
 \else \expandafter \@secondoftwo
 \fi
}%
\providecommand \@ifx [1]{%
 \ifx #1\expandafter \@firstoftwo
 \else \expandafter \@secondoftwo
 \fi
}%
\providecommand \natexlab [1]{#1}%
\providecommand \enquote  [1]{``#1''}%
\providecommand \bibnamefont  [1]{#1}%
\providecommand \bibfnamefont [1]{#1}%
\providecommand \citenamefont [1]{#1}%
\providecommand \href@noop [0]{\@secondoftwo}%
\providecommand \href [0]{\begingroup \@sanitize@url \@href}%
\providecommand \@href[1]{\@@startlink{#1}\@@href}%
\providecommand \@@href[1]{\endgroup#1\@@endlink}%
\providecommand \@sanitize@url [0]{\catcode `\\12\catcode `\$12\catcode
  `\&12\catcode `\#12\catcode `\^12\catcode `\_12\catcode `\%12\relax}%
\providecommand \@@startlink[1]{}%
\providecommand \@@endlink[0]{}%
\providecommand \url  [0]{\begingroup\@sanitize@url \@url }%
\providecommand \@url [1]{\endgroup\@href {#1}{\urlprefix }}%
\providecommand \urlprefix  [0]{URL }%
\providecommand \Eprint [0]{\href }%
\providecommand \doibase [0]{https://doi.org/}%
\providecommand \selectlanguage [0]{\@gobble}%
\providecommand \bibinfo  [0]{\@secondoftwo}%
\providecommand \bibfield  [0]{\@secondoftwo}%
\providecommand \translation [1]{[#1]}%
\providecommand \BibitemOpen [0]{}%
\providecommand \bibitemStop [0]{}%
\providecommand \bibitemNoStop [0]{.\EOS\space}%
\providecommand \EOS [0]{\spacefactor3000\relax}%
\providecommand \BibitemShut  [1]{\csname bibitem#1\endcsname}%
\let\auto@bib@innerbib\@empty
\bibitem [{\citenamefont {Nielsen}\ and\ \citenamefont
  {Ninomiya}(1981{\natexlab{a}})}]{Nielsen:1980rz}%
  \BibitemOpen
  \bibfield  {author} {\bibinfo {author} {\bibfnamefont {H.~B.}\ \bibnamefont
  {Nielsen}}\ and\ \bibinfo {author} {\bibfnamefont {M.}~\bibnamefont
  {Ninomiya}},\ }\bibfield  {title} {\bibinfo {title} {{Absence of Neutrinos on
  a Lattice. 1. Proof by Homotopy Theory}},\ }\href
  {https://doi.org/10.1016/0550-3213(82)90011-6} {\bibfield  {journal}
  {\bibinfo  {journal} {Nucl. Phys. B}\ }\textbf {\bibinfo {volume} {185}},\
  \bibinfo {pages} {20} (\bibinfo {year} {1981}{\natexlab{a}})},\ \bibinfo
  {note} {[Erratum: Nucl.Phys.B 195, 541 (1982)]}\BibitemShut {NoStop}%
\bibitem [{\citenamefont {Nielsen}\ and\ \citenamefont
  {Ninomiya}(1981{\natexlab{b}})}]{Nielsen:1981hk}%
  \BibitemOpen
  \bibfield  {author} {\bibinfo {author} {\bibfnamefont {H.~B.}\ \bibnamefont
  {Nielsen}}\ and\ \bibinfo {author} {\bibfnamefont {M.}~\bibnamefont
  {Ninomiya}},\ }\bibfield  {title} {\bibinfo {title} {{No Go Theorem for
  Regularizing Chiral Fermions}},\ }\href
  {https://doi.org/10.1016/0370-2693(81)91026-1} {\bibfield  {journal}
  {\bibinfo  {journal} {Phys. Lett. B}\ }\textbf {\bibinfo {volume} {105}},\
  \bibinfo {pages} {219} (\bibinfo {year} {1981}{\natexlab{b}})}\BibitemShut
  {NoStop}%
\bibitem [{\citenamefont {Adler}\ and\ \citenamefont
  {Bardeen}(1969)}]{Adler:1969er}%
  \BibitemOpen
  \bibfield  {author} {\bibinfo {author} {\bibfnamefont {S.~L.}\ \bibnamefont
  {Adler}}\ and\ \bibinfo {author} {\bibfnamefont {W.~A.}\ \bibnamefont
  {Bardeen}},\ }\bibfield  {title} {\bibinfo {title} {{Absence of higher order
  corrections in the anomalous axial vector divergence equation}},\ }\href
  {https://doi.org/10.1103/PhysRev.182.1517} {\bibfield  {journal} {\bibinfo
  {journal} {Phys. Rev.}\ }\textbf {\bibinfo {volume} {182}},\ \bibinfo {pages}
  {1517} (\bibinfo {year} {1969})}\BibitemShut {NoStop}%
\bibitem [{\citenamefont {Bell}\ and\ \citenamefont
  {Jackiw}(1969)}]{Bell:1969ts}%
  \BibitemOpen
  \bibfield  {author} {\bibinfo {author} {\bibfnamefont {J.~S.}\ \bibnamefont
  {Bell}}\ and\ \bibinfo {author} {\bibfnamefont {R.}~\bibnamefont {Jackiw}},\
  }\bibfield  {title} {\bibinfo {title} {{A PCAC puzzle: $\pi^0 \to \gamma
  \gamma$ in the $\sigma$ model}},\ }\href {https://doi.org/10.1007/BF02823296}
  {\bibfield  {journal} {\bibinfo  {journal} {Nuovo Cim. A}\ }\textbf {\bibinfo
  {volume} {60}},\ \bibinfo {pages} {47} (\bibinfo {year} {1969})}\BibitemShut
  {NoStop}%
\bibitem [{\citenamefont {Karsten}\ and\ \citenamefont
  {Smit}(1981)}]{Karsten:1980wd}%
  \BibitemOpen
  \bibfield  {author} {\bibinfo {author} {\bibfnamefont {L.~H.}\ \bibnamefont
  {Karsten}}\ and\ \bibinfo {author} {\bibfnamefont {J.}~\bibnamefont {Smit}},\
  }\bibfield  {title} {\bibinfo {title} {{Lattice Fermions: Species Doubling,
  Chiral Invariance, and the Triangle Anomaly}},\ }\href
  {https://doi.org/10.1016/0550-3213(81)90549-6} {\bibfield  {journal}
  {\bibinfo  {journal} {Nucl. Phys. B}\ }\textbf {\bibinfo {volume} {183}},\
  \bibinfo {pages} {103} (\bibinfo {year} {1981})}\BibitemShut {NoStop}%
\bibitem [{\citenamefont {Kogut}\ and\ \citenamefont
  {Susskind}(1975)}]{Kogut:1974ag}%
  \BibitemOpen
  \bibfield  {author} {\bibinfo {author} {\bibfnamefont {J.~B.}\ \bibnamefont
  {Kogut}}\ and\ \bibinfo {author} {\bibfnamefont {L.}~\bibnamefont
  {Susskind}},\ }\bibfield  {title} {\bibinfo {title} {{Hamiltonian Formulation
  of Wilson's Lattice Gauge Theories}},\ }\href
  {https://doi.org/10.1103/PhysRevD.11.395} {\bibfield  {journal} {\bibinfo
  {journal} {Phys. Rev. D}\ }\textbf {\bibinfo {volume} {11}},\ \bibinfo
  {pages} {395} (\bibinfo {year} {1975})}\BibitemShut {NoStop}%
\bibitem [{\citenamefont {Banks}\ \emph {et~al.}(1976)\citenamefont {Banks},
  \citenamefont {Susskind},\ and\ \citenamefont {Kogut}}]{Banks:1975gq}%
  \BibitemOpen
  \bibfield  {author} {\bibinfo {author} {\bibfnamefont {T.}~\bibnamefont
  {Banks}}, \bibinfo {author} {\bibfnamefont {L.}~\bibnamefont {Susskind}},\
  and\ \bibinfo {author} {\bibfnamefont {J.~B.}\ \bibnamefont {Kogut}},\
  }\bibfield  {title} {\bibinfo {title} {{Strong Coupling Calculations of
  Lattice Gauge Theories: (1+1)-Dimensional Exercises}},\ }\href
  {https://doi.org/10.1103/PhysRevD.13.1043} {\bibfield  {journal} {\bibinfo
  {journal} {Phys. Rev. D}\ }\textbf {\bibinfo {volume} {13}},\ \bibinfo
  {pages} {1043} (\bibinfo {year} {1976})}\BibitemShut {NoStop}%
\bibitem [{\citenamefont {Susskind}(1977)}]{Susskind:1976jm}%
  \BibitemOpen
  \bibfield  {author} {\bibinfo {author} {\bibfnamefont {L.}~\bibnamefont
  {Susskind}},\ }\bibfield  {title} {\bibinfo {title} {{Lattice Fermions}},\
  }\href {https://doi.org/10.1103/PhysRevD.16.3031} {\bibfield  {journal}
  {\bibinfo  {journal} {Phys. Rev. D}\ }\textbf {\bibinfo {volume} {16}},\
  \bibinfo {pages} {3031} (\bibinfo {year} {1977})}\BibitemShut {NoStop}%
\bibitem [{\citenamefont {Golterman}\ and\ \citenamefont
  {Smit}(1984)}]{Golterman:1984cy}%
  \BibitemOpen
  \bibfield  {author} {\bibinfo {author} {\bibfnamefont {M.~F.~L.}\
  \bibnamefont {Golterman}}\ and\ \bibinfo {author} {\bibfnamefont
  {J.}~\bibnamefont {Smit}},\ }\bibfield  {title} {\bibinfo {title}
  {{Selfenergy and Flavor Interpretation of Staggered Fermions}},\ }\href
  {https://doi.org/10.1016/0550-3213(84)90424-3} {\bibfield  {journal}
  {\bibinfo  {journal} {Nucl. Phys. B}\ }\textbf {\bibinfo {volume} {245}},\
  \bibinfo {pages} {61} (\bibinfo {year} {1984})}\BibitemShut {NoStop}%
\bibitem [{\citenamefont {Chatterjee}\ \emph {et~al.}(2025)\citenamefont
  {Chatterjee}, \citenamefont {Pace},\ and\ \citenamefont
  {Shao}}]{Chatterjee:2024gje}%
  \BibitemOpen
  \bibfield  {author} {\bibinfo {author} {\bibfnamefont {A.}~\bibnamefont
  {Chatterjee}}, \bibinfo {author} {\bibfnamefont {S.~D.}\ \bibnamefont
  {Pace}},\ and\ \bibinfo {author} {\bibfnamefont {S.-H.}\ \bibnamefont
  {Shao}},\ }\bibfield  {title} {\bibinfo {title} {{Quantized Axial Charge of
  Staggered Fermions and the Chiral Anomaly}},\ }\href
  {https://doi.org/10.1103/PhysRevLett.134.021601} {\bibfield  {journal}
  {\bibinfo  {journal} {Phys. Rev. Lett.}\ }\textbf {\bibinfo {volume} {134}},\
  \bibinfo {pages} {021601} (\bibinfo {year} {2025})},\ \Eprint
  {https://arxiv.org/abs/2409.12220} {arXiv:2409.12220 [hep-th]} \BibitemShut
  {NoStop}%
\bibitem [{\citenamefont {Xu}(2025)}]{Xu:2025hfs}%
  \BibitemOpen
  \bibfield  {author} {\bibinfo {author} {\bibfnamefont {L.-X.}\ \bibnamefont
  {Xu}},\ }\bibfield  {title} {\bibinfo {title} {{Staggered Fermions with
  Chiral Anomaly Cancellation}},\ }\href@noop {} {\  (\bibinfo {year}
  {2025})},\ \Eprint {https://arxiv.org/abs/2501.10837} {arXiv:2501.10837
  [hep-lat]} \BibitemShut {NoStop}%
\bibitem [{\citenamefont {Yamaoka}(2025)}]{Yamaoka:2025sdm}%
  \BibitemOpen
  \bibfield  {author} {\bibinfo {author} {\bibfnamefont {T.}~\bibnamefont
  {Yamaoka}},\ }\bibfield  {title} {\bibinfo {title} {{Quantized Axial Charge
  in the Hamiltonian Approach to Wilson Fermions}},\ }\href@noop {} {\
  (\bibinfo {year} {2025})},\ \Eprint {https://arxiv.org/abs/2504.10263}
  {arXiv:2504.10263 [hep-lat]} \BibitemShut {NoStop}%
\bibitem [{\citenamefont {Wang}\ and\ \citenamefont
  {Wen}(2023)}]{Wang:2013yta}%
  \BibitemOpen
  \bibfield  {author} {\bibinfo {author} {\bibfnamefont {J.}~\bibnamefont
  {Wang}}\ and\ \bibinfo {author} {\bibfnamefont {X.-G.}\ \bibnamefont {Wen}},\
  }\bibfield  {title} {\bibinfo {title} {{Nonperturbative regularization of
  (1+1)-dimensional anomaly-free chiral fermions and bosons: On the equivalence
  of anomaly matching conditions and boundary gapping rules}},\ }\href
  {https://doi.org/10.1103/PhysRevB.107.014311} {\bibfield  {journal} {\bibinfo
   {journal} {Phys. Rev. B}\ }\textbf {\bibinfo {volume} {107}},\ \bibinfo
  {pages} {014311} (\bibinfo {year} {2023})},\ \Eprint
  {https://arxiv.org/abs/1307.7480} {arXiv:1307.7480 [hep-lat]} \BibitemShut
  {NoStop}%
\bibitem [{\citenamefont {Wang}\ and\ \citenamefont
  {Wen}(2018)}]{Wang:2018ugf}%
  \BibitemOpen
  \bibfield  {author} {\bibinfo {author} {\bibfnamefont {J.}~\bibnamefont
  {Wang}}\ and\ \bibinfo {author} {\bibfnamefont {X.-G.}\ \bibnamefont {Wen}},\
  }\bibfield  {title} {\bibinfo {title} {{A Solution to the 1+1D Gauged Chiral
  Fermion Problem}},\ }\href {https://doi.org/10.1103/PhysRevD.99.111501}
  {\bibfield  {journal} {\bibinfo  {journal} {Phys. Rev. D}\ }\textbf {\bibinfo
  {volume} {99}},\ \bibinfo {pages} {111501} (\bibinfo {year} {2018})},\
  \Eprint {https://arxiv.org/abs/1807.05998} {arXiv:1807.05998 [hep-lat]}
  \BibitemShut {NoStop}%
\bibitem [{\citenamefont {Razamat}\ and\ \citenamefont
  {Tong}(2021)}]{Razamat:2020kyf}%
  \BibitemOpen
  \bibfield  {author} {\bibinfo {author} {\bibfnamefont {S.~S.}\ \bibnamefont
  {Razamat}}\ and\ \bibinfo {author} {\bibfnamefont {D.}~\bibnamefont {Tong}},\
  }\bibfield  {title} {\bibinfo {title} {{Gapped Chiral Fermions}},\ }\href
  {https://doi.org/10.1103/PhysRevX.11.011063} {\bibfield  {journal} {\bibinfo
  {journal} {Phys. Rev. X}\ }\textbf {\bibinfo {volume} {11}},\ \bibinfo
  {pages} {011063} (\bibinfo {year} {2021})},\ \Eprint
  {https://arxiv.org/abs/2009.05037} {arXiv:2009.05037 [hep-th]} \BibitemShut
  {NoStop}%
\bibitem [{\citenamefont {Tong}(2022)}]{Tong:2021phe}%
  \BibitemOpen
  \bibfield  {author} {\bibinfo {author} {\bibfnamefont {D.}~\bibnamefont
  {Tong}},\ }\bibfield  {title} {\bibinfo {title} {{Comments on symmetric mass
  generation in 2d and 4d}},\ }\href {https://doi.org/10.1007/JHEP07(2022)001}
  {\bibfield  {journal} {\bibinfo  {journal} {JHEP}\ }\textbf {\bibinfo
  {volume} {07}},\ \bibinfo {pages} {001}},\ \Eprint
  {https://arxiv.org/abs/2104.03997} {arXiv:2104.03997 [hep-th]} \BibitemShut
  {NoStop}%
\bibitem [{\citenamefont {Zeng}\ \emph {et~al.}(2022)\citenamefont {Zeng},
  \citenamefont {Zhu}, \citenamefont {Wang},\ and\ \citenamefont
  {You}}]{Zeng:2022grc}%
  \BibitemOpen
  \bibfield  {author} {\bibinfo {author} {\bibfnamefont {M.}~\bibnamefont
  {Zeng}}, \bibinfo {author} {\bibfnamefont {Z.}~\bibnamefont {Zhu}}, \bibinfo
  {author} {\bibfnamefont {J.}~\bibnamefont {Wang}},\ and\ \bibinfo {author}
  {\bibfnamefont {Y.-Z.}\ \bibnamefont {You}},\ }\bibfield  {title} {\bibinfo
  {title} {{Symmetric Mass Generation in the 1+1 Dimensional Chiral Fermion
  3-4-5-0 Model}},\ }\href {https://doi.org/10.1103/PhysRevLett.128.185301}
  {\bibfield  {journal} {\bibinfo  {journal} {Phys. Rev. Lett.}\ }\textbf
  {\bibinfo {volume} {128}},\ \bibinfo {pages} {185301} (\bibinfo {year}
  {2022})},\ \Eprint {https://arxiv.org/abs/2202.12355} {arXiv:2202.12355
  [cond-mat.str-el]} \BibitemShut {NoStop}%
\bibitem [{\citenamefont {Wang}\ and\ \citenamefont
  {You}(2022)}]{Wang:2022ucy}%
  \BibitemOpen
  \bibfield  {author} {\bibinfo {author} {\bibfnamefont {J.}~\bibnamefont
  {Wang}}\ and\ \bibinfo {author} {\bibfnamefont {Y.-Z.}\ \bibnamefont {You}},\
  }\bibfield  {title} {\bibinfo {title} {{Symmetric Mass Generation}},\ }\href
  {https://doi.org/10.3390/sym14071475} {\bibfield  {journal} {\bibinfo
  {journal} {Symmetry}\ }\textbf {\bibinfo {volume} {14}},\ \bibinfo {pages}
  {1475} (\bibinfo {year} {2022})},\ \Eprint {https://arxiv.org/abs/2204.14271}
  {arXiv:2204.14271 [cond-mat.str-el]} \BibitemShut {NoStop}%
\bibitem [{\citenamefont {Wang}(2022)}]{Wang:2022fzc}%
  \BibitemOpen
  \bibfield  {author} {\bibinfo {author} {\bibfnamefont {J.}~\bibnamefont
  {Wang}},\ }\bibfield  {title} {\bibinfo {title} {{CT or P problem and
  symmetric gapped fermion solution}},\ }\href
  {https://doi.org/10.1103/PhysRevD.106.125007} {\bibfield  {journal} {\bibinfo
   {journal} {Phys. Rev. D}\ }\textbf {\bibinfo {volume} {106}},\ \bibinfo
  {pages} {125007} (\bibinfo {year} {2022})},\ \Eprint
  {https://arxiv.org/abs/2207.14813} {arXiv:2207.14813 [hep-th]} \BibitemShut
  {NoStop}%
\bibitem [{\citenamefont {Lu}\ \emph {et~al.}(2023)\citenamefont {Lu},
  \citenamefont {Zeng}, \citenamefont {Wang},\ and\ \citenamefont
  {You}}]{Lu:2022qtc}%
  \BibitemOpen
  \bibfield  {author} {\bibinfo {author} {\bibfnamefont {D.-C.}\ \bibnamefont
  {Lu}}, \bibinfo {author} {\bibfnamefont {M.}~\bibnamefont {Zeng}}, \bibinfo
  {author} {\bibfnamefont {J.}~\bibnamefont {Wang}},\ and\ \bibinfo {author}
  {\bibfnamefont {Y.-Z.}\ \bibnamefont {You}},\ }\bibfield  {title} {\bibinfo
  {title} {{Fermi surface symmetric mass generation}},\ }\href
  {https://doi.org/10.1103/PhysRevB.107.195133} {\bibfield  {journal} {\bibinfo
   {journal} {Phys. Rev. B}\ }\textbf {\bibinfo {volume} {107}},\ \bibinfo
  {pages} {195133} (\bibinfo {year} {2023})},\ \Eprint
  {https://arxiv.org/abs/2210.16304} {arXiv:2210.16304 [cond-mat.str-el]}
  \BibitemShut {NoStop}%
\bibitem [{\citenamefont {Onogi}\ \emph {et~al.}(2025)\citenamefont {Onogi},
  \citenamefont {Wada},\ and\ \citenamefont {Yamaoka}}]{Onogi:2025tev}%
  \BibitemOpen
  \bibfield  {author} {\bibinfo {author} {\bibfnamefont {T.}~\bibnamefont
  {Onogi}}, \bibinfo {author} {\bibfnamefont {H.}~\bibnamefont {Wada}},\ and\
  \bibinfo {author} {\bibfnamefont {T.}~\bibnamefont {Yamaoka}},\ }\bibfield
  {title} {\bibinfo {title} {{Discrete symmetry and 't Hooft anomalies for 3450
  model}},\ }\href {https://doi.org/10.22323/1.466.0378} {\bibfield  {journal}
  {\bibinfo  {journal} {PoS}\ }\textbf {\bibinfo {volume} {LATTICE2024}},\
  \bibinfo {pages} {378} (\bibinfo {year} {2025})},\ \Eprint
  {https://arxiv.org/abs/2501.18156} {arXiv:2501.18156 [hep-lat]} \BibitemShut
  {NoStop}%
\bibitem [{\citenamefont {Gioia}\ and\ \citenamefont
  {Thorngren}(2025)}]{Gioia:2025bhl}%
  \BibitemOpen
  \bibfield  {author} {\bibinfo {author} {\bibfnamefont {L.}~\bibnamefont
  {Gioia}}\ and\ \bibinfo {author} {\bibfnamefont {R.}~\bibnamefont
  {Thorngren}},\ }\bibfield  {title} {\bibinfo {title} {{Exact Chiral
  Symmetries of 3+1D Hamiltonian Lattice Fermions}},\ }\href@noop {} {\
  (\bibinfo {year} {2025})},\ \Eprint {https://arxiv.org/abs/2503.07708}
  {arXiv:2503.07708 [cond-mat.str-el]} \BibitemShut {NoStop}%
\bibitem [{\citenamefont {Pace}\ \emph {et~al.}(2025)\citenamefont {Pace},
  \citenamefont {Kim}, \citenamefont {Chatterjee},\ and\ \citenamefont
  {Shao}}]{Pace:2025rfu}%
  \BibitemOpen
  \bibfield  {author} {\bibinfo {author} {\bibfnamefont {S.~D.}\ \bibnamefont
  {Pace}}, \bibinfo {author} {\bibfnamefont {M.~L.}\ \bibnamefont {Kim}},
  \bibinfo {author} {\bibfnamefont {A.}~\bibnamefont {Chatterjee}},\ and\
  \bibinfo {author} {\bibfnamefont {S.-H.}\ \bibnamefont {Shao}},\ }\bibfield
  {title} {\bibinfo {title} {{Parity anomaly from LSM: exact valley symmetries
  on the lattice}},\ }\href@noop {} {\  (\bibinfo {year} {2025})},\ \Eprint
  {https://arxiv.org/abs/2505.04684} {arXiv:2505.04684 [cond-mat.str-el]}
  \BibitemShut {NoStop}%
\bibitem [{\citenamefont {Catterall}\ \emph {et~al.}(2025)\citenamefont
  {Catterall}, \citenamefont {Pradhan},\ and\ \citenamefont
  {Samlodia}}]{Catterall:2025vrx}%
  \BibitemOpen
  \bibfield  {author} {\bibinfo {author} {\bibfnamefont {S.}~\bibnamefont
  {Catterall}}, \bibinfo {author} {\bibfnamefont {A.}~\bibnamefont {Pradhan}},\
  and\ \bibinfo {author} {\bibfnamefont {A.}~\bibnamefont {Samlodia}},\
  }\bibfield  {title} {\bibinfo {title} {{Symmetries and Anomalies of
  Hamiltonian Staggered Fermions}},\ }\href@noop {} {\  (\bibinfo {year}
  {2025})},\ \Eprint {https://arxiv.org/abs/2501.10862} {arXiv:2501.10862
  [hep-lat]} \BibitemShut {NoStop}%
\bibitem [{\citenamefont {Li}\ \emph {et~al.}(2024)\citenamefont {Li},
  \citenamefont {Wang},\ and\ \citenamefont {You}}]{Li:2024dpq}%
  \BibitemOpen
  \bibfield  {author} {\bibinfo {author} {\bibfnamefont {Y.-Y.}\ \bibnamefont
  {Li}}, \bibinfo {author} {\bibfnamefont {J.}~\bibnamefont {Wang}},\ and\
  \bibinfo {author} {\bibfnamefont {Y.-Z.}\ \bibnamefont {You}},\ }\bibfield
  {title} {\bibinfo {title} {{Quantum Many-Body Lattice C-R-T Symmetry:
  Fractionalization, Anomaly, and Symmetric Mass Generation}},\ }\href@noop {}
  {\  (\bibinfo {year} {2024})},\ \Eprint {https://arxiv.org/abs/2412.19691}
  {arXiv:2412.19691 [cond-mat.str-el]} \BibitemShut {NoStop}%
\bibitem [{\citenamefont {Banks}\ \emph {et~al.}(1982)\citenamefont {Banks},
  \citenamefont {Dothan},\ and\ \citenamefont {Horn}}]{Banks:1982iq}%
  \BibitemOpen
  \bibfield  {author} {\bibinfo {author} {\bibfnamefont {T.}~\bibnamefont
  {Banks}}, \bibinfo {author} {\bibfnamefont {Y.}~\bibnamefont {Dothan}},\ and\
  \bibinfo {author} {\bibfnamefont {D.}~\bibnamefont {Horn}},\ }\bibfield
  {title} {\bibinfo {title} {{GEOMETRIC FERMIONS}},\ }\href
  {https://doi.org/10.1016/0370-2693(82)90571-8} {\bibfield  {journal}
  {\bibinfo  {journal} {Phys. Lett. B}\ }\textbf {\bibinfo {volume} {117}},\
  \bibinfo {pages} {413} (\bibinfo {year} {1982})}\BibitemShut {NoStop}%
\bibitem [{\citenamefont {Kilcup}\ and\ \citenamefont
  {Sharpe}(1987)}]{Kilcup:1986dg}%
  \BibitemOpen
  \bibfield  {author} {\bibinfo {author} {\bibfnamefont {G.~W.}\ \bibnamefont
  {Kilcup}}\ and\ \bibinfo {author} {\bibfnamefont {S.~R.}\ \bibnamefont
  {Sharpe}},\ }\bibfield  {title} {\bibinfo {title} {{A Tool Kit for Staggered
  Fermions}},\ }\href {https://doi.org/10.1016/0550-3213(87)90285-9} {\bibfield
   {journal} {\bibinfo  {journal} {Nucl. Phys. B}\ }\textbf {\bibinfo {volume}
  {283}},\ \bibinfo {pages} {493} (\bibinfo {year} {1987})}\BibitemShut
  {NoStop}%
\bibitem [{\citenamefont {Catterall}(2021)}]{Catterall:2020fep}%
  \BibitemOpen
  \bibfield  {author} {\bibinfo {author} {\bibfnamefont {S.}~\bibnamefont
  {Catterall}},\ }\bibfield  {title} {\bibinfo {title} {{Chiral lattice
  fermions from staggered fields}},\ }\href
  {https://doi.org/10.1103/PhysRevD.104.014503} {\bibfield  {journal} {\bibinfo
   {journal} {Phys. Rev. D}\ }\textbf {\bibinfo {volume} {104}},\ \bibinfo
  {pages} {014503} (\bibinfo {year} {2021})},\ \Eprint
  {https://arxiv.org/abs/2010.02290} {arXiv:2010.02290 [hep-lat]} \BibitemShut
  {NoStop}%
\bibitem [{\citenamefont {Butt}\ \emph {et~al.}(2021)\citenamefont {Butt},
  \citenamefont {Catterall}, \citenamefont {Pradhan},\ and\ \citenamefont
  {Toga}}]{Butt:2021brl}%
  \BibitemOpen
  \bibfield  {author} {\bibinfo {author} {\bibfnamefont {N.}~\bibnamefont
  {Butt}}, \bibinfo {author} {\bibfnamefont {S.}~\bibnamefont {Catterall}},
  \bibinfo {author} {\bibfnamefont {A.}~\bibnamefont {Pradhan}},\ and\ \bibinfo
  {author} {\bibfnamefont {G.~C.}\ \bibnamefont {Toga}},\ }\bibfield  {title}
  {\bibinfo {title} {{Anomalies and symmetric mass generation for
  K{\"a}hler-Dirac fermions}},\ }\href
  {https://doi.org/10.1103/PhysRevD.104.094504} {\bibfield  {journal} {\bibinfo
   {journal} {Phys. Rev. D}\ }\textbf {\bibinfo {volume} {104}},\ \bibinfo
  {pages} {094504} (\bibinfo {year} {2021})},\ \Eprint
  {https://arxiv.org/abs/2101.01026} {arXiv:2101.01026 [hep-th]} \BibitemShut
  {NoStop}%
\bibitem [{\citenamefont {Catterall}\ and\ \citenamefont
  {Pradhan}(2022)}]{Catterall:2022ukg}%
  \BibitemOpen
  \bibfield  {author} {\bibinfo {author} {\bibfnamefont {S.}~\bibnamefont
  {Catterall}}\ and\ \bibinfo {author} {\bibfnamefont {A.}~\bibnamefont
  {Pradhan}},\ }\bibfield  {title} {\bibinfo {title} {{Induced topological
  gravity and anomaly inflow from K{\"a}hler-Dirac fermions in odd
  dimensions}},\ }\href {https://doi.org/10.1103/PhysRevD.106.014509}
  {\bibfield  {journal} {\bibinfo  {journal} {Phys. Rev. D}\ }\textbf {\bibinfo
  {volume} {106}},\ \bibinfo {pages} {014509} (\bibinfo {year} {2022})},\
  \Eprint {https://arxiv.org/abs/2201.00750} {arXiv:2201.00750 [hep-th]}
  \BibitemShut {NoStop}%
\bibitem [{\citenamefont {Catterall}(2023)}]{Catterall:2022jky}%
  \BibitemOpen
  \bibfield  {author} {\bibinfo {author} {\bibfnamefont {S.}~\bibnamefont
  {Catterall}},\ }\bibfield  {title} {\bibinfo {title} {{\textquoteright{}t
  Hooft anomalies for staggered fermions}},\ }\href
  {https://doi.org/10.1103/PhysRevD.107.014501} {\bibfield  {journal} {\bibinfo
   {journal} {Phys. Rev. D}\ }\textbf {\bibinfo {volume} {107}},\ \bibinfo
  {pages} {014501} (\bibinfo {year} {2023})},\ \Eprint
  {https://arxiv.org/abs/2209.03828} {arXiv:2209.03828 [hep-lat]} \BibitemShut
  {NoStop}%
\bibitem [{\citenamefont {Lieb}\ \emph {et~al.}(1961)\citenamefont {Lieb},
  \citenamefont {Schultz},\ and\ \citenamefont {Mattis}}]{Lieb:1961fr}%
  \BibitemOpen
  \bibfield  {author} {\bibinfo {author} {\bibfnamefont {E.~H.}\ \bibnamefont
  {Lieb}}, \bibinfo {author} {\bibfnamefont {T.}~\bibnamefont {Schultz}},\ and\
  \bibinfo {author} {\bibfnamefont {D.}~\bibnamefont {Mattis}},\ }\bibfield
  {title} {\bibinfo {title} {{Two soluble models of an antiferromagnetic
  chain}},\ }\href {https://doi.org/10.1016/0003-4916(61)90115-4} {\bibfield
  {journal} {\bibinfo  {journal} {Annals Phys.}\ }\textbf {\bibinfo {volume}
  {16}},\ \bibinfo {pages} {407} (\bibinfo {year} {1961})}\BibitemShut
  {NoStop}%
\bibitem [{\citenamefont {Affleck}\ and\ \citenamefont
  {Lieb}(1986)}]{Affleck:1986pq}%
  \BibitemOpen
  \bibfield  {author} {\bibinfo {author} {\bibfnamefont {I.}~\bibnamefont
  {Affleck}}\ and\ \bibinfo {author} {\bibfnamefont {E.~H.}\ \bibnamefont
  {Lieb}},\ }\bibfield  {title} {\bibinfo {title} {{A Proof of Part of
  Haldane's Conjecture on Spin Chains}},\ }\href
  {https://doi.org/10.1007/BF00400304} {\bibfield  {journal} {\bibinfo
  {journal} {Lett. Math. Phys.}\ }\textbf {\bibinfo {volume} {12}},\ \bibinfo
  {pages} {57} (\bibinfo {year} {1986})}\BibitemShut {NoStop}%
\bibitem [{\citenamefont {Oshikawa}(2000)}]{Oshikawa2000TopologicalAT}%
  \BibitemOpen
  \bibfield  {author} {\bibinfo {author} {\bibfnamefont {M.}~\bibnamefont
  {Oshikawa}},\ }\bibfield  {title} {\bibinfo {title} {Topological approach to
  luttinger's theorem and the fermi surface of a kondo lattice},\ }\href
  {https://api.semanticscholar.org/CorpusID:9806160} {\bibfield  {journal}
  {\bibinfo  {journal} {Physical review letters}\ }\textbf {\bibinfo {volume}
  {84 15}},\ \bibinfo {pages} {3370} (\bibinfo {year} {2000})}\BibitemShut
  {NoStop}%
\bibitem [{\citenamefont {Hastings}(2004)}]{Hastings:2003zx}%
  \BibitemOpen
  \bibfield  {author} {\bibinfo {author} {\bibfnamefont {M.~B.}\ \bibnamefont
  {Hastings}},\ }\bibfield  {title} {\bibinfo {title} {{Lieb-Schultz-Mattis in
  higher dimensions}},\ }\href {https://doi.org/10.1103/PhysRevB.69.104431}
  {\bibfield  {journal} {\bibinfo  {journal} {Phys. Rev. B}\ }\textbf {\bibinfo
  {volume} {69}},\ \bibinfo {pages} {104431} (\bibinfo {year} {2004})},\
  \Eprint {https://arxiv.org/abs/cond-mat/0305505} {arXiv:cond-mat/0305505}
  \BibitemShut {NoStop}%
\bibitem [{\citenamefont {Chang}\ \emph {et~al.}(2019)\citenamefont {Chang},
  \citenamefont {Lin}, \citenamefont {Shao}, \citenamefont {Wang},\ and\
  \citenamefont {Yin}}]{Chang:2018iay}%
  \BibitemOpen
  \bibfield  {author} {\bibinfo {author} {\bibfnamefont {C.-M.}\ \bibnamefont
  {Chang}}, \bibinfo {author} {\bibfnamefont {Y.-H.}\ \bibnamefont {Lin}},
  \bibinfo {author} {\bibfnamefont {S.-H.}\ \bibnamefont {Shao}}, \bibinfo
  {author} {\bibfnamefont {Y.}~\bibnamefont {Wang}},\ and\ \bibinfo {author}
  {\bibfnamefont {X.}~\bibnamefont {Yin}},\ }\bibfield  {title} {\bibinfo
  {title} {{Topological Defect Lines and Renormalization Group Flows in Two
  Dimensions}},\ }\href {https://doi.org/10.1007/JHEP01(2019)026} {\bibfield
  {journal} {\bibinfo  {journal} {JHEP}\ }\textbf {\bibinfo {volume} {01}},\
  \bibinfo {pages} {026}},\ \Eprint {https://arxiv.org/abs/1802.04445}
  {arXiv:1802.04445 [hep-th]} \BibitemShut {NoStop}%
\bibitem [{\citenamefont {Wen}(2019)}]{Wen:2018zux}%
  \BibitemOpen
  \bibfield  {author} {\bibinfo {author} {\bibfnamefont {X.-G.}\ \bibnamefont
  {Wen}},\ }\bibfield  {title} {\bibinfo {title} {{Emergent anomalous higher
  symmetries from topological order and from dynamical electromagnetic field in
  condensed matter systems}},\ }\href
  {https://doi.org/10.1103/PhysRevB.99.205139} {\bibfield  {journal} {\bibinfo
  {journal} {Phys. Rev. B}\ }\textbf {\bibinfo {volume} {99}},\ \bibinfo
  {pages} {205139} (\bibinfo {year} {2019})},\ \Eprint
  {https://arxiv.org/abs/1812.02517} {arXiv:1812.02517 [cond-mat.str-el]}
  \BibitemShut {NoStop}%
\bibitem [{\citenamefont {Thorngren}\ and\ \citenamefont
  {Wang}(2024)}]{Thorngren:2019iar}%
  \BibitemOpen
  \bibfield  {author} {\bibinfo {author} {\bibfnamefont {R.}~\bibnamefont
  {Thorngren}}\ and\ \bibinfo {author} {\bibfnamefont {Y.}~\bibnamefont
  {Wang}},\ }\bibfield  {title} {\bibinfo {title} {{Fusion category symmetry.
  Part I. Anomaly in-flow and gapped phases}},\ }\href
  {https://doi.org/10.1007/JHEP04(2024)132} {\bibfield  {journal} {\bibinfo
  {journal} {JHEP}\ }\textbf {\bibinfo {volume} {04}},\ \bibinfo {pages}
  {132}},\ \Eprint {https://arxiv.org/abs/1912.02817} {arXiv:1912.02817
  [hep-th]} \BibitemShut {NoStop}%
\bibitem [{\citenamefont {Tu}\ \emph {et~al.}(2025)\citenamefont {Tu},
  \citenamefont {Long},\ and\ \citenamefont {Else}}]{Tu:2025bqf}%
  \BibitemOpen
  \bibfield  {author} {\bibinfo {author} {\bibfnamefont {Y.-T.}\ \bibnamefont
  {Tu}}, \bibinfo {author} {\bibfnamefont {D.~M.}\ \bibnamefont {Long}},\ and\
  \bibinfo {author} {\bibfnamefont {D.~V.}\ \bibnamefont {Else}},\ }\bibfield
  {title} {\bibinfo {title} {{Anomalies of global symmetries on the lattice}},\
  }\href@noop {} {\  (\bibinfo {year} {2025})},\ \Eprint
  {https://arxiv.org/abs/2507.21209} {arXiv:2507.21209 [cond-mat.str-el]}
  \BibitemShut {NoStop}%
\bibitem [{\citenamefont {Onsager}(1944)}]{Onsager:1943jn}%
  \BibitemOpen
  \bibfield  {author} {\bibinfo {author} {\bibfnamefont {L.}~\bibnamefont
  {Onsager}},\ }\bibfield  {title} {\bibinfo {title} {{Crystal statistics. 1. A
  Two-dimensional model with an order disorder transition}},\ }\href
  {https://doi.org/10.1103/PhysRev.65.117} {\bibfield  {journal} {\bibinfo
  {journal} {Phys. Rev.}\ }\textbf {\bibinfo {volume} {65}},\ \bibinfo {pages}
  {117} (\bibinfo {year} {1944})}\BibitemShut {NoStop}%
\bibitem [{\citenamefont {Choi}\ \emph {et~al.}(2022)\citenamefont {Choi},
  \citenamefont {Cordova}, \citenamefont {Hsin}, \citenamefont {Lam},\ and\
  \citenamefont {Shao}}]{Choi:2021kmx}%
  \BibitemOpen
  \bibfield  {author} {\bibinfo {author} {\bibfnamefont {Y.}~\bibnamefont
  {Choi}}, \bibinfo {author} {\bibfnamefont {C.}~\bibnamefont {Cordova}},
  \bibinfo {author} {\bibfnamefont {P.-S.}\ \bibnamefont {Hsin}}, \bibinfo
  {author} {\bibfnamefont {H.~T.}\ \bibnamefont {Lam}},\ and\ \bibinfo {author}
  {\bibfnamefont {S.-H.}\ \bibnamefont {Shao}},\ }\bibfield  {title} {\bibinfo
  {title} {{Noninvertible duality defects in 3+1 dimensions}},\ }\href
  {https://doi.org/10.1103/PhysRevD.105.125016} {\bibfield  {journal} {\bibinfo
   {journal} {Phys. Rev. D}\ }\textbf {\bibinfo {volume} {105}},\ \bibinfo
  {pages} {125016} (\bibinfo {year} {2022})},\ \Eprint
  {https://arxiv.org/abs/2111.01139} {arXiv:2111.01139 [hep-th]} \BibitemShut
  {NoStop}%
\end{thebibliography}%

\end{document}